\documentclass[12pt]{article} 
\usepackage[sectionbib]{natbib}
\usepackage{array,epsfig,fancyheadings,rotating}
\usepackage[]{hyperref}  
\usepackage{sectsty, secdot}
\sectionfont{\fontsize{12}{14pt plus.8pt minus .6pt}\selectfont}
\renewcommand{\theequation}{\thesection\arabic{equation}}
\subsectionfont{\fontsize{12}{14pt plus.8pt minus .6pt}\selectfont}

\usepackage{amsmath}
\usepackage{amssymb}
\usepackage{amsfonts}
\usepackage{multirow}
\usepackage{amsthm}

\usepackage{mathtools}
\usepackage{booktabs}
\usepackage{enumitem}
\usepackage{array}
\usepackage{bbm}

\usepackage{times}
\usepackage{bm}
\usepackage{multirow}
\usepackage{graphicx}
\usepackage{caption}
\usepackage[titletoc]{appendix}
\usepackage[plain,noend]{algorithm2e}
\usepackage{url} 
\newtheorem{proposition}{Proposition}
\theoremstyle{definition}

\begin{document}


\renewcommand{\baselinestretch}{2}

\renewcommand{\thefootnote}{}
$\ $\par


\fontsize{12}{14pt plus.8pt minus .6pt}\selectfont \vspace{0.8pc}
\centerline{\large\bf A Clustered Gaussian Process Model for Computer Experiments}
\vspace{.4cm} \centerline{Chih-Li Sung$^1$, Benjamin Haaland$^2$, Youngdeok Hwang$^3$, Siyuan Lu$^4$} \vspace{0cm} {\center{\small\it
$^1$Michigan State University,\quad $^2$University of Utah\\$^3$City University of New York,\quad $^4$IBM Thomas J. Watson Research Center\\}}
\vspace{.3cm} \fontsize{9}{11.5pt plus.8pt minus
.6pt}\selectfont

\begin{quotation}
\noindent {\it Abstract:}
A Gaussian process has been one of the important approaches for emulating computer simulations. However, the stationarity assumption for a Gaussian process and the intractability for large-scale dataset limit its availability in practice. In this article, we propose a clustered Gaussian process model which segments the input data into multiple clusters, in each of which a Gaussian process model is performed. The stochastic expectation-maximization is employed to efficiently fit the model. In our simulations as well as a real application to solar irradiance emulation, our proposed method had smaller mean square errors than its main competitors, with competitive computation time, and provides valuable insights from data by discovering the clusters. An R package for the proposed methodology is provided in an open repository.

\vspace{9pt}
\end{quotation}\par

\def\thefigure{\arabic{figure}}
\def\thetable{\arabic{table}}

\renewcommand{\theequation}{\thesection.\arabic{equation}}

\fontsize{12}{14pt plus.8pt minus .6pt}\selectfont

\setcounter{section}{0} 
\setcounter{equation}{0} 

\lhead[\footnotesize\thepage\fancyplain{}\leftmark]{}\rhead[]{\fancyplain{}\rightmark\footnotesize\thepage}

\section{Introduction}
\begin{figure}[ht]
\centering
\includegraphics[width=0.45\linewidth]{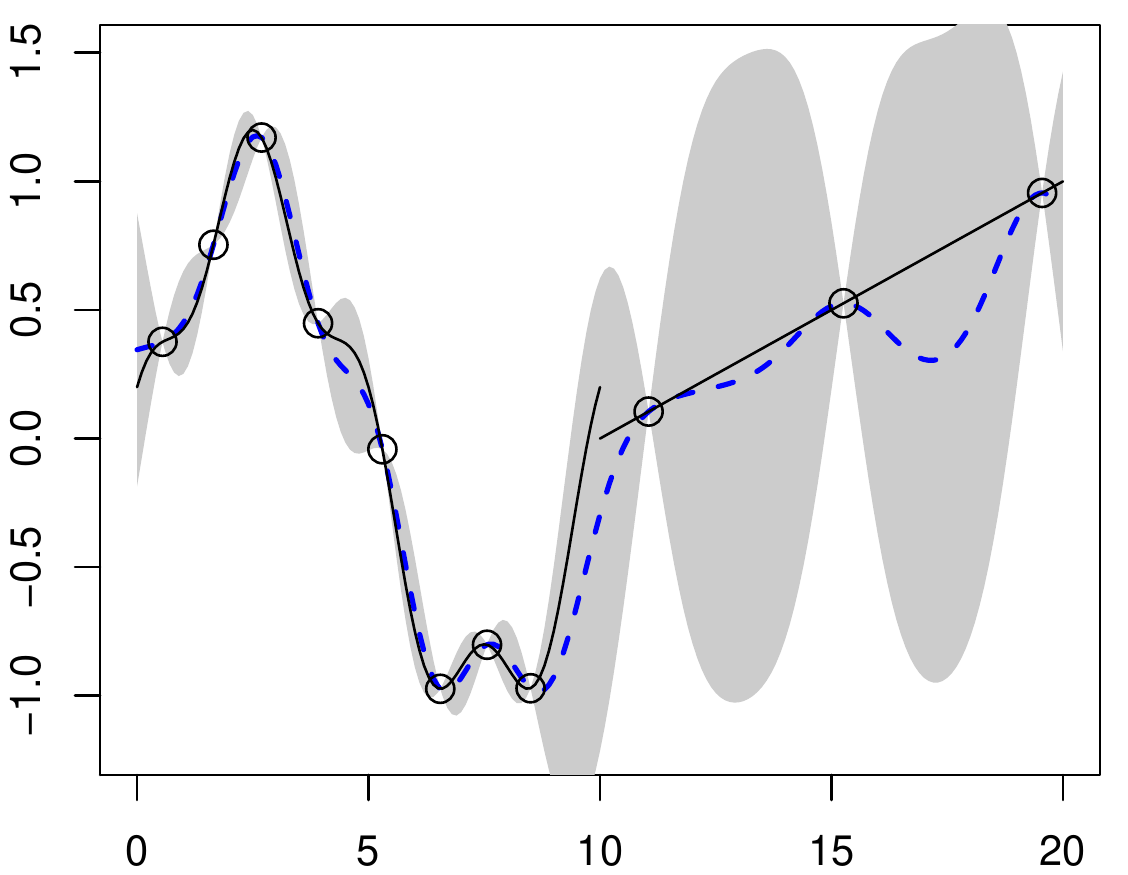}
\caption{An example of stationary Gaussian processes. Black line is the true function, and black dots represent the collected data. The blue dashed line represents a stationary Gaussian process, with the gray shaded region providing a pointwise 95\% confidence band.}
\label{fig:example_introduction}
\end{figure}

A Gaussian process (GP) has been one of the most popular modeling tools in various research topics, such as spatial statistics \citep{stein2012interpolation}, computer experiments \citep{fang2005design,santner2013design}, machine learning \citep{rasmussen2006gaussian}, and robot control \citep{nguyen2011model}. Gaussian processes provide the flexibility for a prior probability distribution over functions in Bayesian inference, and its posterior is not only able to estimate the functional for an unseen point but also has uncertainty information. This explicit probabilistic formulation for GP has proved to be powerful for general function learning problems. However, its use is often limited due to the following challenges. First, GP posterior involves $O(N^3)$ computational complexity and $O(N^2)$ storage where $N$ is the sample size, so that it becomes infeasible for a moderately large data sets, say $N=10^3$. Second, a GP model often considers a stationary covariance function, in the sense that the outputs with the same separation of any two inputs are assumed to have an equal covariance. We call it a stationary GP in the article. This assumption is violated in many practical applications, particularly for nonstationary processes. Figure \ref{fig:example_introduction} demonstrates an illustrative example in \cite{gramacy2009adaptive} where a stationary GP may perform very poorly when the underlying function indeed consists of two different functions: a relatively rough function in the region $x\in[0,10]$ and a simple linear function in the region $x\in[10,20]$. Figure \ref{fig:example_introduction} shows that a stationary GP results in very poor prediction particularly in the region $x\in[10,20]$ with very high uncertainty. See more examples in \cite{higdon1999non,paciorek2006spatial,bui2012adaptive}.

These two challenges for GP have attracted lots of attention lately. To name a few, sparse approximation \citep{quinonero2005unifying,sang2012full}, covariance tapering \citep{furrer2006covariance}, inducing inputs \citep{snelson2006sparse,titsias2009variational}, multi-step interpolation \citep{haaland2011accurate}, special design \citep{plumlee2014fast}, multi-resolution approximation \citep{nychka2014multi}, address the computational issue for large datasets. For nonstationarity, \cite{higdon1999non,paciorek2006spatial,plumlee2017lifted} adopted nonstationary covariance functions for Gaussian processes. \cite{tresp2001mixtures,rasmussen2002infinite,kim2005analyzing,gramacy2008bayesian} considered multiple Gaussian processes by segmentation in the input spaces. \cite{ba2012composite} proposed a composite of two Gaussian processes, which respectively capture a smooth global trend and local details. However, only few of them are able to tackle the nonstationarity and computational issues simultaneously. Exceptions include the multi-resolution functional ANOVA approximation \citep{sung2017multi} which uses group lasso algorithm to identify important basis functions, and the local Gaussian process approximation which selects a small subsample to fit a Gaussian process model for each predictive location \citep{gramacy2015local}.

In this article, we propose a clustered Gaussian process (clustered GP) to address the two challenges simultaneously. The clustered GP makes use of the idea of input segmentation by considering multiple Gaussian processes in the input spaces with a hard-assignment clustering approach, which makes the computation more tractable for large datasets, and also provides some valuable insights about the underlying aspects of data set by showing some grouping structure. 

The remainder of this article is organized as follows. In Section \ref{sec:clusteredGP}, the clustered GP model is introduced with its relationship to the existing methods. In Section \ref{sec:inference}, our estimation and prediction to fit the clustered GP model using a stochastic expectation-maximization algorithm is described. Some computational details are discussed in Section \ref{sec:computationdetails}. In Section \ref{sec:numericalstudy}, some synthetic examples are demonstrated to show the tractability and the prediction performance of the proposed method. A real data application for predicting solar irradiance over the United States is presented in Section \ref{sec:realdata}. Some potential future work is discussed in Section \ref{sec:discussion}. 

\section{Clustered Gaussian Process} \label{sec:clusteredGP}

\subsection{Preliminary: Gaussian Processes}
A brief review for Gaussian processes is first given in this section. A Gaussian process (GP) is a stochastic process whose finite dimensional distributions are defined via a mean function $\mu(x)$ and a covariance function $\Sigma(x,x')$ for $d$-dimensional $x,x'\in\chi\subseteq\mathbb{R}^d$. 
If the function $y(\cdot)$ is a draw from a GP, then we write
\[
y(\cdot)\sim\mathcal{GP}(\mu(\cdot),\Sigma(\cdot,\cdot)).\]
In particular, given $n$ inputs $X=(x_1,\ldots,x_n)$, if $y(\cdot)$ is a GP, then the outputs $Y=(y(x_1),\ldots,y(x_n))$ have a multivariate normal distribution
\[
Y|X\sim\mathcal{N}(\mu(X), \Sigma(X,X)),
\]
where $\mu(X)\in\mathbb{R}^n$ and $\Sigma(X,X)\in\mathbb{R}^{n\times n}$ are defined as $(\mu(X))_i=\mu(x_i)$ and $(\Sigma(X,X))_{i,j}=\Sigma(x_i,x_j)$, respectively. Conventionally, $\mu(\cdot)$ is often assumed to be a constant mean, i.e., $\mu(\cdot)=\mu$, and $\Sigma(\cdot,\cdot)$ is assumed to have the form $\sigma^2\Phi_{\gamma}(\cdot,\cdot)$, where $\Phi_{\gamma}$ is a correlation function with $\Phi_{\gamma}(x,x)=1$ for any $x\in\chi$ and contains the unknown parameter $\gamma$. In addition, $\Phi_{\gamma}$ is often assumed to depend on the displacement between two input locations, that is, $\Phi_{\gamma}(x,x')=R(x-x')$ for some positive-definite function $R$. 
Such a correlation function is called \textit{stationary} correlation function which implies the process $y(\cdot)$ is stationary, since $y(x_1),\ldots,y(x_L)$ and $y(x_1+h),\ldots,y(x_L+h)$ have the same distribution for any $h\in\mathbb{R}^d$ and $x_1,\ldots,x_L,x_1+h,\ldots,x_L+h\in\chi$. A common choice for $\Phi$ is a power correlation function 
\begin{equation}\label{eq:powercov}
\Phi_{\gamma}(x,x')=\exp\{-\|\gamma^T(x-x')\|^p\},
\end{equation}
where $p$ is often fixed to control the smoothness of the output surface, and $\gamma=(\gamma_1,\ldots,\gamma_d)^T$ controls the decay of correlation with respect to the distance between $x$ and $x'$.
Hence, the parameters include $\mu(\cdot),\sigma^2$ and $\gamma$ and can be estimated by either maximum likelihood estimation or Bayesian estimation. See \cite{fang2005design}, \cite{rasmussen2006gaussian} and \cite{santner2013design} for more details. Importantly, when the interest is in the prediction at an untried $x_{\rm{new}}$, whose response could be denoted as $y_{\rm{new}}$, the predictive distribution of $y_{\rm{new}}$ can be derived by the conditional multivariate normal distribution. In particular, one can show that $y_{\rm{new}}|Y,X,x_{\rm{new}}\sim\mathcal{N}(\mu^*,(\sigma^*)^2)$, where 
\begin{equation}\label{eq:predictivemean}
\mu^*=\mu(x_{\rm{new}})+\Phi_{\gamma}(x_{\rm{new}},X)\Phi_{\gamma}(X,X)^{-1}(Y-\mu(X)),
\end{equation}
and 
\begin{equation}\label{eq:predictivevar}
(\sigma^*)^2=\sigma^2\left(1-\Phi_{\gamma}(x_{\rm{new}},X)\Phi_{\gamma}(X,X)^{-1}\Phi_{\gamma}(X,x_{\rm{new}})\right).
\end{equation}
In practice, the unknown parameters $\mu(\cdot),\sigma^2$ and $\gamma$ in \eqref{eq:predictivemean} and \eqref{eq:predictivevar} are replaced by their estimates.

\subsection{Clustered Gaussian Process}
In practice, we might expect the unknown function that we are trying to approximate to exhibit some degree of non-stationarity.
A natural conceptual model to take into account such circumstance would be a mixture GP, where each component of the mixture acts as an approximately stationary model with a high accuracy for a subset of the data. That is,
\begin{align}
y(\cdot)\mid z(\cdot)=k&\sim\mathcal{GP}(\mu_k(\cdot),\sigma_k^2\Phi_{\gamma_k}(\cdot,\cdot)),\quad k=1,\ldots,K,\nonumber\\
{\rm Pr}(z(x)=k) &= g_k(x;\varphi_k),\quad k=1,\ldots,K,\label{eq:modelz}
\end{align}
where $\mu_k(\cdot),\sigma_k^2$ and $\Phi_{\gamma_k}$ are the mean function, variance, and stationary correlation function of the $k$-th GP, and 
$g_k(x,\varphi_k)$ is the probability that $z(x)=k$ with unknown parameter $\varphi_k$ satisfying $\sum^K_{k=1}g_k(x;\varphi_k)=1$ for any $x$. 
It can be seen that in this model, $z(\cdot)$ takes the role of a latent function, which assigns $y(\cdot)$ to one of the $K$ GPs. These assignments introduce a non-stationarity to the response $y(\cdot)$, even though each GP is stationary.

Now, a little notation is  introduced. Given $n$ inputs $X=(x_1,\ldots,x_n)$, denote the corresponding outputs as $Y=(Y(x_1),\ldots,Y(x_n))$.
For cluster $k=1,\ldots,K$, let $\mathcal{P}_k=\{i:z(x_i)=k\}$ denote the set of indices of the observations in cluster $k$. 
Additionally, let $Y_{\mathcal{P}_k}$ and $X_{\mathcal{P}_k}$ respectively denote the (ordered) responses and input locations for the observations from cluster $k$. Then, given $Z=(z_1,\ldots,z_n)\equiv(z(x_1),\ldots,z(x_n))$, the output $Y_{\mathcal{P}_k}$ in each cluster $k$ has the multivariate normal distribution
\begin{equation}\label{eq:likelihoodincluster}
Y_{\mathcal{P}_k}|X_{\mathcal{P}_k} \sim\mathcal{N}(\mu_k(X_{\mathcal{P}_k}), \sigma^2_k \Phi_{\gamma_k}(X_{\mathcal{P}_k},X_{\mathcal{P}_k})),
\end{equation}
where the observed $y_i$'s depend on the response values and locations of the other cluster members, in addition to their corresponding input location $x_i$ within each cluster. 
The latent cluster/mixture component assignments $z_i$ is assumed to be independent across observations $i$ but dependent on input location $x_i$, so that the (unobserved) cluster assignment likelihood is given by
\begin{align}
f(Z|X)&=\text{Pr}(z(x_1)=z_1,\ldots,z(x_n)=z_n)\nonumber\\
&=\prod^n_{i=1} g_{z_i}(x_i;\varphi_{z_i})=\prod^K_{k=1}\prod_{i\in\mathcal{P}_k} g_{k}(x_i;\varphi_{k}).\label{eq:zlikelihood}
\end{align}
Then, by combining \eqref{eq:likelihoodincluster} and \eqref{eq:zlikelihood}, the likelihood function of complete data is 
\begin{align}
f(Y,Z|X)=&f(Y|X,Z)f(Z|X)\nonumber\\
=&\left(\prod_{k=1}^Kf_k(Y_{\mathcal{P}_k}|X_{\mathcal{P}_k};\theta_k)\right)\left(\prod_{k=1}^K\prod_{i\in\mathcal{P}_k}g_k(x_i;\varphi_k)\right)\label{eq:completeDataLik},
\end{align}
where $f_k$ is the probability density function of a multivariate normal distribution with parameters $\theta_k\equiv\{\mu_k(\cdot),\sigma_k^2,\gamma_k\}$.

The clustered GP in \eqref{eq:modelz} is closely related to some of existing methods. If $z(\cdot)$ is a Bayesian treed models, the model becomes close to the Bayesian treed GP of \cite{gramacy2008bayesian}. If $z(\cdot)$ assigns cluster memberships based on a Voronoi tessellation, the model bears some similarity to the model of \cite{kim2005analyzing}. When $z(\cdot)$ is assumed to be a Dirichlet process or a generalized GP, the model becomes similar to the mixtures of GPs of \cite{tresp2001mixtures} and \cite{rasmussen2002infinite}, respectively.  
Despite the similarity, their application is limited in large-scale data setting due to their costly MCMC sampling. 
Some other work, such as \cite{nguyen2009local,zhang2019learning}, chose the assignment based on traditional unsupervised clustering methods, such as $K$-means clustering.

Our modeling approach belongs to the popular model based clustering approach using latent variables within Expectation-Maximization (EM) framework \citep[e.g.,][]{fraley2002model}. A likelihood-based EM approach to estimate the unknown parameters is, however, not straightforward, because strong dependencies among observations due to the GP correlation structure makes computation difficult. One may want to compute the cluster probability $f(Z|X,Y)$, whether for implementing the E-step in the EM algorithm (soft assignment), or updating cluster membership in a $K$-means type algorithm (hard assignment). Unfortunately, the cluster probability $f(Z|X,Y)$ do not factor beyond being proportional to (\ref{eq:completeDataLik}), so we cannot compute the cluster membership for each observation separately from one another even though we assumed that $z_i$ is independent of each other. In the next section, we propose a stochastic EM algorithm to address this issue, along with computational details associated with our approach. 



\section{Statistical Inference via Stochastic EM Algorithm} \label{sec:inference}
In this section, we present our estimation and prediction approach for the model in \eqref{eq:modelz}. Our proposed method addresses the aforementioned challenges using the utilizing stochastic EM algorithm \citep[SEM,][]{celeux1985sem}. SEM algorithm is particularly suitable for our challenges as it leads to a computationally efficient algorithm in clustered GP while avoiding insignificant local maxima of likelihood functions.

\subsection{Stochastic E-step}\label{sec:Estep}
In the EM-algorithm, the E-step computes the expected value of the log posterior of complete data given the observed data $Y$:
\begin{equation}\label{eq:Estepposteriror}
\mathbb{E}[\log f(Y,Z|X)|X,Y,\boldsymbol{\theta},\boldsymbol{\varphi}] + \log\pi(\boldsymbol{\theta}) + \log\pi(\boldsymbol{\varphi}),
\end{equation}
where $\boldsymbol{\theta}=\{\theta_k\}^K_{k=1}$, $\boldsymbol{\varphi}=\{\varphi_k\}^K_{k=1}$, and $\pi(\boldsymbol{\theta})$ and $\pi(\boldsymbol{\varphi})$ are priors of $\boldsymbol{\theta}$ and $\boldsymbol{\varphi}$. We assume $\theta_k$ and $\varphi_k$ are mutually independent through $k=1,\ldots,K$ so 
\begin{equation}\label{eq:priorassumption}
    \log\pi(\boldsymbol{\theta})=\sum^K_{k=1}\log\pi(\theta_k) \quad\text{and}\quad\log\pi(\boldsymbol{\varphi})=\sum^K_{k=1}\log\pi(\varphi_k).
\end{equation}
Computing the expected value requires the cluster probabilities $f(Z|X,Y)$, which cannot be explicitly evaluated. Instead, we adopt a Gibbs sampling, or iterative stochastic hard assignment. The key quantity for this approach is the cluster membership probability for observation $i$ given the data $X,Y$ and the other cluster memberships $Z_{-i}$,
\begin{gather}
\begin{split}
&f(z_i=k|X,Y,Z_{-i})\propto f(Y|X,Z_{-i},z_i=k)f(z_i=k|X,Z_{-i})\\
&=\left(f_k(Y_{\mathcal{P}_k\cup\{i\}}|X_{\mathcal{P}_k\cup\{i\}};\theta_k)\prod_{j\ne k}f_{j}(Y_{\mathcal{P}_{j}\setminus\{i\}}|X_{\mathcal{P}_{j}\setminus\{i\}};\theta_j)\right)g_k(x_i;\varphi_k).\label{GibbsAssignment}
\end{split}
\end{gather}
Despite our highly dependent situation, \eqref{GibbsAssignment} can be calculated in a simple form as shown in Proposition \ref{prop1}. The proof is deferred to Supplementary Material \ref{append:proof}.
\begin{proposition}\label{prop1}
Under the complete data likelihood given in \eqref{eq:completeDataLik},
\begin{equation}\label{eq:proportionprob}
f(z_i=k|X,Y,Z_{-i})\propto \phi((y_i-\mu^*_k)/\sigma_k^*)g_k(x_i;\varphi_k),\quad \text{where}
\end{equation}
\begin{gather}
\begin{split}
&\mu^*_k=\mu_k(x_i)+\Phi_{\gamma_k}(x_i,X_{\mathcal{P}_k\setminus\{i\}})\Phi_{\gamma_k}(X_{\mathcal{P}_k\setminus\{i\}},X_{\mathcal{P}_k\setminus\{i\}})^{-1}\left(Y_{\mathcal{P}_k\setminus\{i\}}-\mu_k(X_{\mathcal{P}_k\setminus\{i\}})\right),\\
&(\sigma_k^*)^2=\sigma_k^2\left(1-\Phi_{\gamma_k}(x_i,X_{\mathcal{P}_k\setminus\{i\}})\Phi_{\gamma_k}(X_{\mathcal{P}_k\setminus\{i\}},X_{\mathcal{P}_k\setminus\{i\}})^{-1}\Phi_{\gamma_k}(X_{\mathcal{P}_k\setminus\{i\}},x_i)\right),\label{eq:GibbsAssignment2}
\end{split}
\end{gather}
where $\phi$ is the density probability function of a standard normal distribution.
\end{proposition}

Proposition \ref{prop1} implies the cluster is assigned very intuitively. For an unknown predictive location $x_i$, the predictive distribution of each cluster $k$ is a normal distribution with mean $\mu_{k}^{*}$ and variance $(\sigma_k^*)^2$ as in \eqref{eq:predictivemean} and \eqref{eq:predictivevar}. Thus, the membership of $z_i$ then can be determined from the probability density function of cluster $k$ at $y_i$, and the probability mass function $g_k$ of membership $k$ at $x_i$. The membership is likely to be assigned to $k$th class if (a)
if $y_i$ is closer to $\mu_k^{*}$ with regard to the scale $\sigma_k^{*}$; (b) $g_k$ has a high mass probability at location $x_i$.


Once \eqref{eq:proportionprob} is available for each $i$ and $k$, a random cluster assignment can be drawn from a multinomial distribution. Each step of this Gibbs scheme satisfies detailed balance (assuming none of the probabilities/densities in (\ref{eq:proportionprob}) equal zero), so eventually this process produces samples from $f(Z|X,Y)$. Hence, the cluster membership samples can be used to approximate 
quantities depending on $f(Z|X,Y)$, such as the expectation in \eqref{eq:Estepposteriror}. Further, partitioned matrix inverse and determinant formulas \citep{harville1998matrix} allow one to update the augmented and diminished Gaussian densities in $O(n_k^2)$ time, where $n_k$ is the number of observations in cluster $k$. The details are provided in Supplementary Material \ref{app:updatematrixinverse}. In total, each iteration going through all the observations would take at most $O(\sum_{k=1}^K n_k^3)$. One may ease computational burden by controlling the maximum number of observations in each cluster, denoted by $n_{\max}$, then the computation becomes $O(Kn_{\max}^3)$ in total. Computation in this step can be easily distributed over multiple cores, in particular, \eqref{eq:GibbsAssignment2} can be done separately for different $k$. The detailed algorithm is given in Stochastic E-step of Supplementary Material \ref{alg:clusteringalgorithm}.



\subsection{M-step}\label{seq:Mstep}
Once a random assignment drawn from  $\tilde{\mathcal{P}}_k=\{i:\tilde{z}_i=k\}$ is available from the stochastic E-step, we can proceed to the M-step. Let $\tilde{Z}$ denote the random assignment, and $\tilde{\mathcal{P}}_k=\{i:\tilde{z}_i=k\}$ the set of indices of the observations in cluster $k$ assigned in $\tilde{Z}$, respectively.
From \eqref{eq:completeDataLik} and \eqref{eq:priorassumption}, the log posterior of complete data in  \eqref{eq:Estepposteriror} is approximately by 
\begin{align*}
&\log f(Y,\tilde{Z}|X,\boldsymbol{\theta},\boldsymbol{\varphi})+\log\pi(\boldsymbol{\theta}) + \log\pi(\boldsymbol{\varphi})\\
=&\sum_{k=1}^K \log f_k(Y_{\tilde{\mathcal{P}}_k}|X_{\tilde{\mathcal{P}}_k};\theta_k)+\sum_{k=1}^K\sum_{i\in\tilde{\mathcal{P}}_k}\log g_k(x_i;\varphi_k)+\sum^K_{k=1}\log\pi(\theta_k)+\sum^K_{k=1}\log\pi(\varphi_k).
\end{align*}
The maximum a posteriori probability (MAP) estimate $\{\hat{\theta}_k\}^K_{k=1}$ and $\{\hat{\varphi}_k\}^K_{k=1}$ can then be obtained by maximizing 
$$\sum_{k=1}^K \log \left(f_k(Y_{\tilde{\mathcal{P}}_k}|X_{\tilde{\mathcal{P}}_k};\theta_k)\pi(\theta_k)\right)\quad \text{and}\quad \sum_{k=1}^K\left(\sum_{i\in\tilde{\mathcal{P}}_k}\log g_k(x_i;\varphi_k)+\log\pi(\varphi_k)\right),$$ respectively. In particular, $\sum_{k=1}^K \log \left(f_k(Y_{\tilde{\mathcal{P}}_k}|X_{\tilde{\mathcal{P}}_k};\theta_k)\pi(\theta_k)\right)$ can be optimized by maximizing each component $f_k(Y_{\tilde{\mathcal{P}}_k}|X_{\tilde{\mathcal{P}}_k};\theta_k)\pi(\theta_k)$, which is proportional to the posterior distribution of the $k$-th GP. 
The choice for the prior of $\theta_k$ and its resulting posterior can be found in Chapters 3 and 4 of \cite{santner2013design}. The computation for M-step can be done for $K$ clusters separately, which can be efficiently parallelized as in  Supplementary Material \ref{alg:clusteringalgorithm}. 

\subsection{Prediction}\label{sec:prediction}
Predicting the responses $y_{\rm new}$ at a new input location $x_{\rm new}$ can be challenging, since the cluster assignment $z_{\rm new}$ at the new location is unknown. Given the assignment $\tilde{Z}=(\tilde{z}(x_1),\ldots,\tilde{z}(x_n))$ and the estimates $\{\hat{\theta}_k,\hat{\varphi}_k\}^K_{k=1}$ returned in the SEM algorithm, we perform the predictive distribution of $y_{\rm new}$ by weighted averaging across the clustered GPs:
\begin{align*}
f(y_{\rm new}|x_{\rm new},X,Y,\tilde{Z})=&\sum^K_{k=1}f(y_{\rm new}|z_{\rm new}=k,x_{\rm new},X,Y,\tilde{Z})f(z_{\rm new}=k|x_{\rm new},X,Y,\tilde{Z})\\
=&\sum^K_{k=1}\phi((y_{\rm new}-\hat{\mu}^*_k)/\hat{\sigma}^*_k)g_k(x_{\rm new};\hat{\varphi}_k),
\end{align*}
where 
\begin{align*}
\hat{\mu}^*_k&=\hat{\mu}_k(x_{\rm new})+\Phi_{\hat{\gamma}_k}(x_{\rm new},X_{\tilde{\mathcal{P}}_k})\Phi_{\hat{\gamma}_k}(X_{\tilde{\mathcal{P}}_k},X_{\tilde{\mathcal{P}}_k})^{-1}\left(Y_{\tilde{\mathcal{P}}_k}-\hat{\mu}_k(X_{\tilde{\mathcal{P}}_k})\right),\\
(\hat{\sigma}^*_k)^2&=\hat{\sigma}_k^2\left(1-\Phi_{\hat{\gamma}_k}(x_{\rm new},X_{\tilde{\mathcal{P}}_k})\Phi_{\hat{\gamma}_k}(X_{\tilde{\mathcal{P}}_k},X_{\tilde{\mathcal{P}}_k})^{-1}\Phi_{\hat{\gamma}_k}(X_{\tilde{\mathcal{P}}_k},x_{\rm new})\right).
\end{align*}
Thus, the prediction mean of $y_{\rm new}$ is
\begin{equation}\label{eq:clusteredpredictionmean}
\hat{y}_{\rm new}:=\mathbb{E}[y_{\rm new}|x_{\rm new},X,Y,\tilde{Z}]=\sum^K_{k=1}\hat{\mu}^*_kg_k(x_{\rm new};\hat{\varphi}_k),
\end{equation}
with its variance
\begin{align*}
\mathbb{V}[y_{\rm new}|x_{\rm new},X,Y,\tilde{Z}]=&\mathbb{E}[\mathbb{V}[y_{\rm new}|z_{\rm new},x_{\rm new},X,Y,\tilde{Z}]]+\mathbb{V}[\mathbb{E}[y_{\rm new}|z_{\rm new},x_{\rm new},X,Y,\tilde{Z}]]\\
=&\sum^K_{k=1}(\hat{\sigma}^*_k)^2g_k(x_{\rm new};\hat{\varphi}_k)+\sum^K_{k=1}(\hat{\mu}^*_k)^2g_k(x_{\rm new};\hat{\varphi}_k)-\left(\sum^K_{k=1}\hat{\mu}^*_kg_k(x_{\rm new};\hat{\varphi}_k)\right)^2.
\end{align*}
The $q$-th quantile of $y_{\rm new}$, which will be used for constructing confidence intervals, has no closed form but can be calculated by finding the value of $y$ for which $\int^y_{-\infty}f(t|x_{\rm new},X,Y,\tilde{Z}){\rm d}t=q$, which is equivalent to solving
\[
\sum^K_{k=1}\left(\int^y_{-\infty}\phi((t-\hat{\mu}^*_k)/\hat{\sigma}^*_k){\rm d}t\right)g_k(x_{\rm new};\hat{\varphi}_k)=q.
\]
The summation and integration are interchangeable because the probability density function is finite. The equation can be solved numerically, for example, using a line search or generating Monte Carlo samples.

\section{Computational details}\label{sec:computationdetails}
In this section, we provide some computational details for the proposed SEM that we have provided in Section \ref{sec:inference}. In particular, we discuss the possible choices in each element in the algorithm, with the focus on the specific implementation that we have adopted.

\subsection{Choices for class assignment model}\label{sec:gfunction}
The model for $z(\cdot)$ in \eqref{eq:modelz} determines the latent class distribution of the cluster assignment, where $g_k$ is the conditional probability that $z(x)=k$ given an input $x$. Amongst several possibilities to model $z(\cdot)$, one can consider a $K$-class multinomial logistic regression,
\[
{\rm Pr}(z(x)=k)=g_k(x;\varphi_k)=\frac{\exp\{\beta_{0,k}+\beta^T_k x\}}{\sum^{K}_{j=1}\exp\{\beta_{0,j}+\beta^T_j x\}},
\]
for $k=1,\ldots,K-1$ and ${\rm Pr}(z(x)=K)=1-\sum^{K-1}_{j=1}{\rm Pr}(z(x)=j)$, where $\beta_{0,k}$ is the intercept, $\beta_k$ is a $d$-dimensional coefficient of $x$, and $\varphi_k=(\beta_1,\ldots,\beta_{K-1})$. Alternatively, one can also consider the linear discriminant analysis (LDA) or quadratic discriminant analysis (QDA) methods by assuming
\[
g_k(x;\varphi_k)=\phi(x;\nu_k,\Sigma_k) \quad\text{for}\quad k = 1,\ldots,K,
\]
where $\phi(x;\nu_k,\Sigma_k)$ is the density probability function of a (multivariate) normal distribution with mean $\nu_k$ and covariance $\Sigma_k$. LDA assumes $\Sigma_1=\ldots=\Sigma_K$, while QDA assumes the covariances can be different. The multinomial logistic regression and LDA methods are closely connected, which often result in similar linear decision boundaries of the $K$ classes. QDA methods, on the other hand, result in quadratic decision boundaries. From our preliminary investigation, the clustered Gaussian processes with these models give similar prediction results. As such, we only present $K$-class multinomial logistic regression hereinafter.


\subsection{Initialization}\label{sec:initialization}
The SEM algorithm can be sensitive to the initialization. One may run many initializations and select the one that gives the optimal criterion. This is, however, computational intensive especially for large data sets. One potential initialization is the $K$-means clusters or other unsupervised clustering algorithms solely based on the input $X$. This initialization enables the clustered GP to make the input locations of each cluster close to each other and distant from the ones of other clusters, which often leads to a nice model interpretation.
Although this initialization may end up with a local optimum, the cluster structure can help model perform well by efficiently exchanging the class assignment. Our preliminary investigation showed that the clustered GPs based on the initialization often result in promising prediction accuracy along with nice model interpretation. In Sections \ref{sec:numericalstudy} and \ref{sec:realdata}, the initialization of $K$-means clusters will be used.

\subsection{Stopping criteria}\label{sec:stoppingrule}
The iteration in the SEM algorithm in  Supplementary Material \ref{alg:clusteringalgorithm} needs a stopping criterion to determine a convergence. 
For this purpose, we propose to use leave-one-out cross-validation (LOOCV), so that the algorithm stops when the cross-validated prediction error does not improve. LOOCV iteratively holds out one particular location, trains on the data remaining at other locations, and then makes prediction for the held-out location. Although LOOCV is often too expensive to implement in many situations as the model has to fit $n$ times in each iteration, the clustered GP has an efficient shortcut that makes the LOOCV very affordable. Specifically, denote $\tilde{y}_i$ as the prediction mean based on all data except $i$-th observation and $y_i$ as the real output of $i$-th observation, then based on \eqref{eq:clusteredpredictionmean},
\begin{equation*}
    \tilde{y}_i=\sum^K_{k=1}\hat{\mu}^{(-i)}_kg_k(x_i;\hat{\varphi}_k),
\end{equation*}
where 
\begin{equation}\label{eq:loocvmean}
\hat{\mu}^{(-i)}_k=\hat{\mu}_k(x_i)+\Phi_{\hat{\gamma}_k}(x_i,X_{\tilde{\mathcal{P}}_k\setminus\{i\}})\Phi_{\hat{\gamma}_k}(X_{\tilde{\mathcal{P}}_k\setminus\{i\}},X_{\tilde{\mathcal{P}}_k\setminus\{i\}})^{-1}\left(Y_{\tilde{\mathcal{P}}_k\setminus\{i\}}-\hat{\mu}_k(X_{\tilde{\mathcal{P}}_k\setminus\{i\}})\right).
\end{equation}
For those $i$s which do not belong to $\tilde{\mathcal{P}}_k$, \eqref{eq:loocvmean} becomes 
\[
\hat{\mu}^{(-i)}_k=\hat{\mu}_k(x_i)+\Phi_{\hat{\gamma}_k}(x_i,X_{\tilde{\mathcal{P}}_k})\Phi_{\hat{\gamma}_k}(X_{\tilde{\mathcal{P}}_k},X_{\tilde{\mathcal{P}}_k})^{-1}\left(Y_{\tilde{\mathcal{P}}_k}-\hat{\mu}_k(X_{\tilde{\mathcal{P}}_k})\right),
\]
and for those $i$s which belong to $\tilde{\mathcal{P}}_k$, \eqref{eq:loocvmean} can be simplified to 
\begin{equation} \label{eqn:simplified-loocv}
\hat{\mu}^{(-i)}_k=\hat{\mu}_k(x_i)-\frac{1}{q_{ii}}\sum^{n_k}_{j\neq i}q_{ij}(y_j-\hat{\mu}_k(x_j)),    
\end{equation}
where $q_{ij}$ is the $(i,j)$-th element of $\Phi_{\hat{\gamma}_k}(X_{\tilde{\mathcal{P}}_k},X_{\tilde{\mathcal{P}}_k})^{-1}$. Then, the LOOCV root-mean-squared error (RMSE) is
\begin{equation*}
    \sqrt{\frac{1}{n}\sum^n_{i=1}(y_i-\tilde{y}_i)^2}= \sqrt{\frac{1}{n}\sum^n_{i=1}\left(y_i-\sum^K_{k=1}\hat{\mu}^{(-i)}_kg_k(x_i;\hat{\varphi}_k)\right)^2}.
\end{equation*}
This computation costs at most $O(Kn_
{\rm max}^3)$, which is same as the SEM algorithm.

\subsection{The choice of $K$}
The number of clusters $K$ plays an important role for the degree of non-stationarity of approximation functions which may affect approximation accuracy. A natural choice is using cross-validation with different $K$'s to target a small prediction error, such as the LOOCV RMSE described in Section \ref{sec:stoppingrule}. Other choices using bootstrap techniques to estimate prediction error also can be considered, such as the 632+ bootstrap method of \cite{efron1997improvements}. \cite{kohavi1995study} which explicitly discussed the comparison between cross-validation and bootstrap from bias and variance point of view and comprehensive numerical experiments were conducted therein. 
For the purpose of saving computational cost, we choose the $K$ that gives the lowest LOOCV RMSE, because LOOCV RMSE can be computed efficiently for clustered GPs as given in \eqref{eqn:simplified-loocv}. 

\subsection{Remarks on the alternative implementation}

The SEM and prediction can be modified in a more fully Bayesian fashion using the Monte Carlo samples from the posterior distribution of $\{z(x_i)\}^n_{i=1}$, $\{\theta_k,\varphi_k\}^K_{k=1}$ with a Gibbs routine to generate predictions. The computational burden for this direction, however, can be prohibitively heavy in a large-data context. In particular, saving samples from the posteriors requires enormous amounts of storage for large data sets. Using the returned assignment $\tilde{Z}$ and the MAPs $\{\hat{\theta}_k,\hat{\varphi}_k\}^K_{k=1}$ can be an efficient alternative with representative samples for more efficient fitting and prediction procedures.

The MAP estimation in the M-step can be replaced with maximum likelihood (ML) estimation, or simply by letting the prior distributions of $\{\theta_k\}^K_{k=1}$ and $\{\varphi_k\}^K_{k=1}$ be uniform. 
Under some regularity conditions, the ML estimators $\{\hat{\theta}_k\}^K_{k=1}$ and $\{\hat{\varphi}_k\}^K_{k=1}$ can be shown to have an asymptotically normal distribution in such approach. We refer the asymptotic properties of the parameter inference to \cite{nielsen2000stochastic}.

\section{Numerical study}\label{sec:numericalstudy}
In this section we present several exemplar functions to demonstrate the effectiveness of clustered Gaussian processes. 
We first present examples with lower dimensional inputs to visually present the cluster structure and the benefit from non-stationary modeling and then to an example with higher-dimension inputs.
Throughout, the $K$-means clusters are chosen as the initialization, and the $K$-class multinomial logistic regression is modeled for $z(\cdot)$. The iteration in the SEM algorithm stops when LOOCV does not improve or the number of iterations exceeds the preset maximum. We select the assignment $\tilde{Z}$ which results in the lowest LOOCV RMSE during the iterations, which will be illustrated in Section \ref{sec:twodimexample}.
Power correlation function of \eqref{eq:powercov} with $p=2$ is chosen. 
Both of the mean functions $\mu(\cdot)$ and $\mu_k(\cdot)$ of the stationary GP and the clustered GP are assumed to be constant. 
For each cluster, a small nugget, $10^{-6}$, is added when fitting a GP model for numerical stability. In addition, we let the prior distributions of $\{\theta_k\}^K_{k=1}$ and $\{\varphi_k\}^K_{k=1}$ be uniform.

\subsection{One-dimensional synthetic data}\label{sec:onedimexample}

\begin{figure}[h!]
\centering
\includegraphics[width=0.8\linewidth]{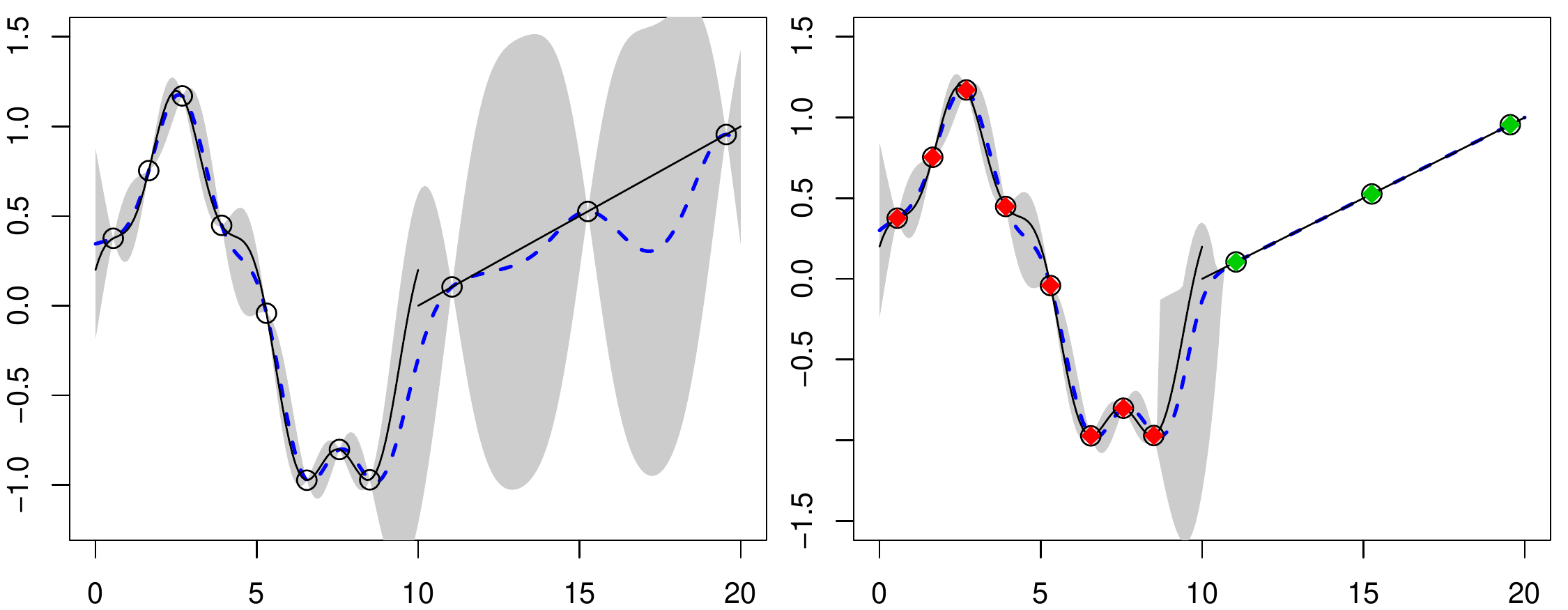}
\caption{One-dimensional synthetic data. The left and right panels illustrate predictors by a stationary GP and a clustered GP, respectively. Black line is the true function, black circles are input locations, and blue dotted lines are the predictors, with the gray shaded region providing a pointwise 95\% confidence band. Red, green, and blue dots in the right panels represent different clusters.}
\label{fig:example_synthetic2}
\end{figure}

Consider an example from \cite{gramacy2009adaptive}, which is a modification to the example in \cite{higdon2002space}. Suppose that the true function is 
\begin{equation*}
  f(x)=\begin{cases}
    \sin(0.2\pi x)+0.2\cos(0.8\pi x), & \text{if $x<10$}.\\
    0.1 x-1, & \text{otherwise}
  \end{cases}
\end{equation*}
and 11 unequally spaced points from $[0,20]$ are chosen. The black lines in the top panels of Figure \ref{fig:example_synthetic2} demonstrate this function, and it can be seen that the function is discontinuous at $x=10$. When the data are fitted by a stationary GP, it can be seen in the top-left panel of Figure \ref{fig:example_synthetic2} that the prediction within region $[10,20]$ performs very badly with large uncertainty. \cite{ba2012composite} explained that the constant mean assumption for GP is violated so the predictor tends to revert to the global mean, whose estimate is 0.208 by maximum likelihood estimation in this example. This consequence is frequently observed especially at the locations far away from input locations. Moreover, the constant variance assumption for GP is also violated. The function in the region $[0,10]$ is rougher than that in the region $[10,20]$. Therefore, the variance estimate for region $[10,20]$ tends to be inflated by averaging with that of region $[0,10]$, which leads to the erratic prediction in this region. On the other hand, clustered GP introduces some degree of non-stationarity by considering a mixture GP, which is shown in the top-right panel of Figure \ref{fig:example_synthetic2}. Two subsets of the data are represented as red and green dots, which are given by the assignment $\tilde{Z}$ returned in the SEM algorithm, and both are fitted by stationary GPs. The mean estimates of the GPs are -0.045 and 0.529, respectively. It can be seen that the predictor performs much better than a stationary GP, especially at the locations within region $[10,20]$, in terms of prediction accuracy and uncertainty quantification. The most uncertain region is located on the boundary of two clusters, which is expected because the assignment of cluster membership is more uncertain in the region. One potential remedy of improving the accuracy on the boundaries will be discussed in Section \ref{sec:discussion}. 

Two more one-dimensional synthetic data generated from the exemplar functions of \cite{xiong2007non} and \cite{montagna2016computer} are presented in Supplementary Material \ref{supp:1d_example}, in which both examples show that the clustered GP yields better prediction accuracy than a stationary GP.

\subsection{Two-dimensional synthetic data}\label{sec:twodimexample}
\begin{figure}[h!]
\centering
\includegraphics[width=\linewidth]{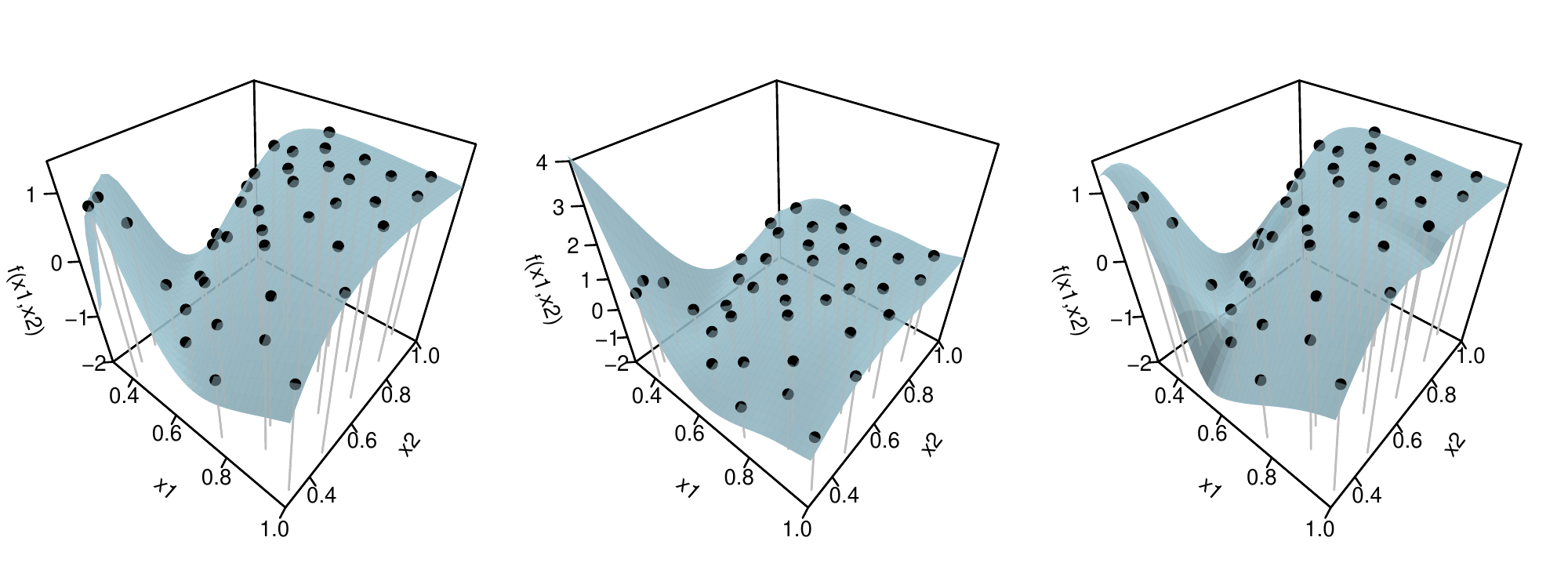}
\caption{Two-dimensional wavy function, and the input locations which are shown as black dots. The left panel is the true wavy function, the middle panel is the predictive surface of a stationary GP, and the right panel is the predictive surface of a clustered GP.}
\label{fig:wavyfunction}
\end{figure}

In this section, the selection of $K$ and the stopping rule using LOOCV RMSE will be demonstrated. Consider a wavy function, which also appeared in \cite{ba2012composite} and \cite{montagna2016computer}. The wavy function is
\[
f(x_1,x_2)=\sin\left(\frac{1}{x_1x_2}\right),
\]
where $x_1,x_2\in[0.3,1]$. The function is illustrated in the left panel of Figure \ref{fig:wavyfunction}, in which it fluctuates rapidly when $x_1$ and $x_2$ are small and gets smoother as they increase toward 1. A 40-run maximin distance Latin hypercube design \citep{morris1995exploratory} from $[0.3,1]^2$ is chosen to select the input locations at which the wavy function is evaluated. These locations are shown as black dots. A stationary GP and a clustered GP with $K=3$ are performed on these locations, whose predictive surfaces are shown in the middle and right panels of Figure \ref{fig:wavyfunction}. It can be seen that the stationary GP performs fairly poorly as $x_1$ and $x_2$ are small, while the clustered GP generally has better prediction performance over the input space. To evaluate the prediction performance quantitatively, we predict the responses at 1296 ($=36\times 36$) equally spaced points from $[0.3,1]^2$ as the test points, and compute their RMSEs, that is,
\[
\left(\frac{1}{n_{\rm test}}\sum^{n_{\rm test}}_{i=1}\left(f(x_1,x_2)-\hat{f}(x_1,x_2)\right)^2\right)^{1/2},
\]
where $n_{\rm test}$ is the number of test points and  $\hat{f}(x_1,x_2)$ is the predicted value at $x_1$ and $x_2$. In this example, the clustered GP outperforms the stationary GP in terms of prediction accuracy, where their RMSEs are 0.1872 and 0.3569, respectively.

\begin{figure}[h!]
\centering
\includegraphics[width=0.9\linewidth]{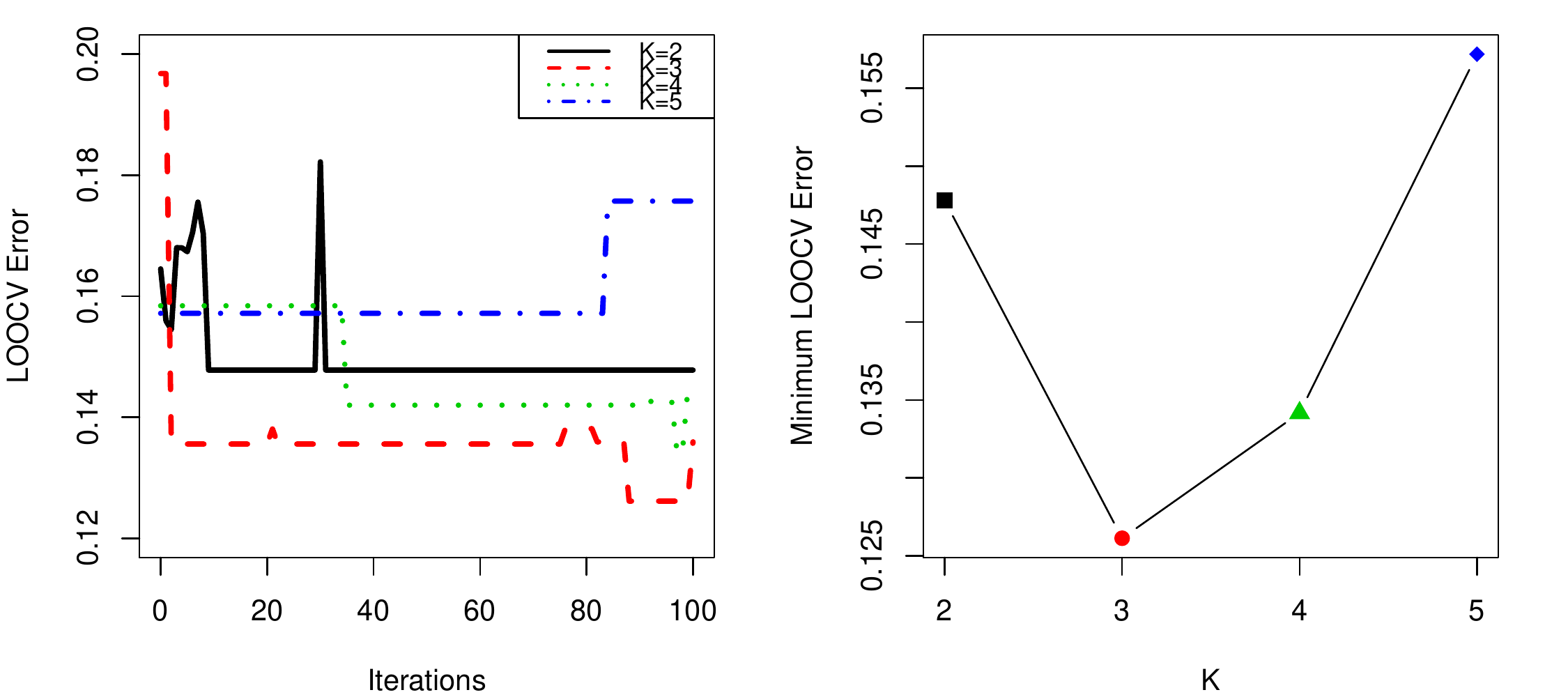}
\caption{The LOOCV RMSEs with $K=2,3,4$ and $5$ during the 100 iteration of the SEM algorithm (left), and the minimum LOOCV RMSE of each choice of $K$ (right).}
\label{fig:loocv}
\end{figure}
\begin{figure}[h!]
\centering
\includegraphics[width=\linewidth]{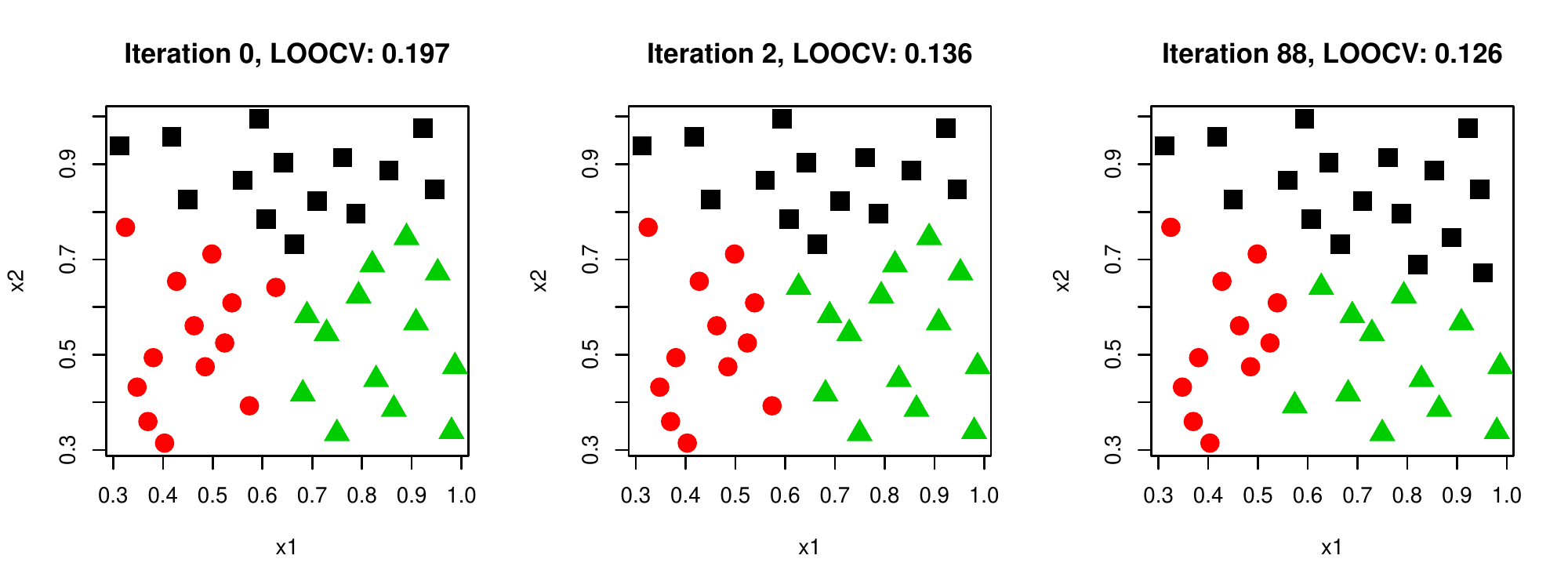}
\caption{The cluster assignments at iteration 0, 2, and 88 of the SEM algorithm and their LOOCV RMSEs.}
\label{fig:iteration}
\end{figure}

Figure \ref{fig:loocv} demonstrates the stopping rule and the selection of $K$ discussed in Section \ref{sec:computationdetails}. The left panel presents the LOOCV RMSEs of $K=2,3,4$ and $5$ during the 100 iterations of the SEM algorithm. It shows that even though the LOOCV RMSE of initial iteration of $K=3$ is larger than other choices of $K$, the error drops rapidly and ends up with a lower LOOCV error at 88-th iteration. For each choice of $K$, we chose the assignment of the iteration that results in the minimum LOOCV RMSE as the final assignment $\tilde{Z}$ for prediction. The right panel presents the minimum LOOCV RMSE of each choice of $K$ in the 100 iterations, and it shows that $K=3$ gives the lowest LOOCV RMSE so it was selected in this example. Figure \ref{fig:iteration} demonstrates the assignments at iteration 0, 2, and 88 when $K=3$. The assignment at iteration 0 represents initial assignment, which is the $K$-means clusters as described in Section \ref{sec:initialization}, whose LOOCV RMSE is 0.197. The LOOCV RMSE then drops dramatically in the second iteration from 0.197 to 0.136 with only one assignment switched, that is, the point $x_1=0.627, x_2=0.641$ is from circle to triangle cluster. With more iterations and more assignments switched, the LOOCV error decreases to 0.126 at iteration 88. The final assignment gives an intuitive explanation: the points when both of $x_1$ and $x_2$ are small, where the true function has a sharp change, appear to belong to the same cluster (see the circle cluster).

\subsection{Borehole function}\label{sec:borehole}
In the section, a borehole function, a more complex exemplar function with 8-dimensional input, is considered to examine the scalability of clustered GP. The borehole function models water flow through a borehole, and has been commonly used for testing a wide variety of methods in computer experiments because of its quick evaluation. The borehole function is given by 
\begin{equation}\label{eq:borehole}
f(x)=\frac{2\pi T_u(H_u-H_l)}{\ln(r/r_w)\left(1+\frac{2LT_u}{\ln(r/r_w)r_w^2K_w}+\frac{T_u}{T_l}\right) },
\end{equation}
where $r_w\in[0.05, 0.15]$ is the radius of borehole, $r\in[100, 50000]$ is the radius of influence, $T_u\in[63070, 115600]$ is the transmissivity of upper aquifer, $H_u\in[990, 1110]$ is the potentiometric head of upper aquifer, $T_l\in[63.1, 116]$ is the transmissivity of lower aquifer, $H_l\in[700, 820]$ is the potentiometric head of lower aquifer, $L\in[1120, 1680]$ is the length of borehole, and $K_w\in[9855, 12045]$ is the hydraulic conductivity of borehole. 

Consider $n$ uniformly distributed input locations in the input space described above and $n_{\rm test}=10,000$ random input locations in the same input space for examining prediction accuracy, whose outputs are evaluated from \eqref{eq:borehole}.
Four methods are compared, including a stationary GP, local GP \citep{gramacy2015local}, multi-resolution functional ANOVA (MRFA) \citep{sung2017multi}, and clustered GP. These methods are implemented using \texttt{R} \citep{R2015} via packages \texttt{mlegp} \citep{mlegp}, \texttt{laGP} \citep{gramacy2015lagp}, \texttt{MRFA} \citep{sung2017mrfa}, \texttt{clusterGP}, on a MacBook Pro laptop with 2.6 GHz Intel Core i7 and 16GB of RAM. 
For the purpose of demonstration, $K=n/200$ was chosen for all the cases.
For \texttt{laGP}, \texttt{MRFA} and \texttt{clusterGP}, 10 CPU cores were requested via \texttt{foreach} \citep{pkgforeach} for parallel computing.

Table \ref{tab:borehole} shows the performance of the four methods, 
in terms of computation time and prediction accuracy.
It can be seen that the stationary GP is only feasible when $n=1,000$, while other three methods are feasible for larger $n$. Even when a stationary GP is feasible, the accuracy is worse than \texttt{MRFA} and \texttt{clusterGP}. Among the four methods, \texttt{clusterGP} has better accuracy with reasonable computation time. \texttt{MRFA} has slightly larger predictive errors with faster computation. On the other hand, local GP has larger predictive errors, even though the computation is faster. One may consider a different setting for local GP (e.g., the size of subsample) which may lead to better accuracy.

\section{Solar irradiance prediction}\label{sec:realdata}
Predicting solar irradiance, or the power per unit area produced by electromagnetic radiation, plays a very important role in power balancing and determining the viability of potential sites for harvesting solar power. One dataset can be brought to bear on this problem is the simulations from the North American Mesoscale Forecast System (NAM) \citep{rogers2009ncep}, which is one of the major weather models run by the National Centers for Environmental Prediction (NCEP) for producing weather forecasts. We extract the solar irradiance (global horizontal irradiance) simulations from the NAM model at the locations of 1,535 Remote Automatic Weather Station (RAWS) \citep{zachariassen2003review} sites in the contiguous United States. Note that the RAWS stations are not uniformly distributed. 
Figure \ref{fig:solarstation} visualizes the available locations and their corresponding solar irradiance with the average taken over one year, which can be seen that many promising locations for solar farms are sparsely covered particularly in the Midwest. These locations of interest are considered for solar energy forecasting. Detail description of the dataset can be found in \cite{hwang2018} and \cite{sun2018synthesizing}. Similar to \cite{sun2018synthesizing}, here we work with average irradiance values over one year from the NAM simulations for each of 1535 spatial locations (as shown in Figure \ref{fig:solarstation}), and the research interest of this study is making accurate prediction for solar irradiance at those unavailable locations. 

In Figure \ref{fig:solarstation}, it appears that some relatively high solar irradiance are measured compared to their neighborhood, such as at the location on the coordinate $(-93.57, 45.99)$, and some relatively low solar irradiance are measured such as at the location on the coordinate $(-93.16, 33.69)$. These instances may suggest that heterogeneity rather than homogeneity in the input-output relationships should be considered. The assumption of identical covariance function throughout the input domain for stationary GPs, therefore, is likely to fail and may result in poor performance, as shown in the examples of Section \ref{sec:numericalstudy}.

A clustered GP is performed on this dataset, where similar setup in Section \ref{sec:twodimexample} was used. 
We first use the LOOCV to determine the number of clusters $K$. The left panel of Figure \ref{fig:determineK} shows the LOOCV RMSEs of $K=15,25,35,45$ during 20 iterations of the SEM algorithm, and the right panel shows the minimum LOOCV RMSEs with respect to different choices of $K$. Based on the right panel, it appears that $K=35$ has the lowest LOOCV RMSE among $K=10, 15, 20, 25, 30, 35, 40, 45, 50$, which suggests that $K=35$ is a good choice for predicting solar irradiance. Similar to the numerical study in Section \ref{sec:numericalstudy}, we chose the assignment of the iteration which results in the lowest LOOCV RMSE as the final assignment $\tilde{Z}$.
The assignment $\tilde{Z}$ is visualized in Figure \ref{fig:clustering}, where the 35 clusters are presented as different colors and numbers. It appears that
the assignments for the clusters are flexible that do not rely on linear decision boundaries. For example, cluster 26 are mostly located on Michigan and part of Pennsylvania and New York, which tells us that some common aspects of the solar irradiance are shared in those areas adjacent to Great Lakes, even though they are not spatially connected. The example shows that the clustering can provide a useful insight for discovering groups and identifying interesting insight of a dataset.


To examine its prediction accuracy, we use LOOCV RMSEs as the prediction error and compare with a recent emulation method in \cite{sun2017emulating},
where they proposed a multi-resolution global/local GP emulation by extending the idea of local GP \citep{gramacy2015local}, 
and their latter work in \cite{sun2018synthesizing} applied this method to the same NAM simulation data herein. \cite{sun2018synthesizing} reported the LOOCV errors of the multi-resolution global/local GP emulation as well as the ordinary stationary GP. The results together with our proposed method are presented in Figure \ref{fig:predictioncomparison}. The figure presents the true solar irradiance (top left) and the LOOCV predictions of the stationary GP (top right), the multi-resolution global/local GP (bottom left), and the clustered GP with $K=35$ (bottom right), along with their corresponding LOOCV RMSEs in the titles.  It can be seen that, the stationary GP does a poor job in predicting the solar irradiance, the LOOCV predictions of which are all essentially equal which implies that almost all of the pattern remains in the errors, which in turn gives a high LOOCV RMSE (23.20). Performances of the multi-resolution global/local GP as well as the clustered GP on the other hand are very good, the result of which may suggest that the nonstationarity should be taken into account for this dataset. Although the LOOCV predictions are visually similar, the LOOCV RMSE of the clustered GP is slightly lower than the multi-resolution global/local GP (9.11 and 9.74, respectively). In particular, it appears that the clustered GP has better prediction accuracy in the Northeast and Southeast, whereas the multi-resolution global/local GP tends to be more smooth over the whole space.

\section{Discussion}\label{sec:discussion}
In this paper, we proposed a clustered Gaussian process which entertains computational advantages and tackles the nonstationarity limitations of stationary Gaussian processes. Unlike traditional clustering methods in an unsupervised way, the clusters in the clustered GP are \textit{supervised} by the response - that is, it makes use of the response in order to partition the input domain that not only clusters the observations that have similar features, but also that have the same stationary process in the response. This clustering algorithm is implemented using a stochastic EM algorithm, which is available in an open repository. Examples including the application of solar irradiance simulations show that the method not only has advantages in computation and prediction accuracy, but also enables discovery of interesting insights by interpreting the clusters.

The clustered GP indicates several avenues for future research. First, the stochastic EM algorithm can be modified in an online fashion. That is, if the data is available in a sequential order, then instead of starting from the new dataset augmented with the additional data, the algorithm can be modified to update the clusters and the best predictor for future data at each step. For example, the solar irradiance simulations are available in every hour, so the modified algorithm can be used to update the clusters and predict future data in real time, which can save substantial computational cost and storage especially when the training sample size is extremely large. 
Moreover, to reduce the prediction uncertainty on the boundary between two regions (see, for example, $x=10$ in Figure \ref{fig:example_synthetic2}), it is conceivable to apply the idea of ``patchwork'' in \cite{park2018patchwork} by patching the GPs on the boundary, which can mitigate the discontinuous problem that may degrade the prediction accuracy. We leave these to our future work.

\vspace{5mm}
\noindent \textbf{Supplementary Materials}: The online supplementary materials contain the detailed proof of Proposition \ref{prop1}, the detailed SEM algorithm in Section \ref{sec:inference}, supporting tables and figures for Sections \ref{sec:numericalstudy} and \ref{sec:realdata}. An \texttt{R} package \texttt{GPcluster} for implementing the proposed method is available in an open repository.

\vspace{5mm}

{
\bibliography{paper-ref}

\begin{thebibliography}{}

\bibitem[Ba and Joseph, 2012]{ba2012composite}
Ba, S. and Joseph, V.~R. (2012).
\newblock Composite {G}aussian process models for emulating expensive
  functions.
\newblock {\em The Annals of Applied Statistics}, 6(4):1838--1860.

\bibitem[Bui-Thanh et~al., 2012]{bui2012adaptive}
Bui-Thanh, T., Ghattas, O., and Higdon, D. (2012).
\newblock Adaptive hessian-based nonstationary gaussian process response
  surface method for probability density approximation with application to
  bayesian solution of large-scale inverse problems.
\newblock {\em SIAM Journal on Scientific Computing}, 34(6):A2837--A2871.

\bibitem[Celeux and Diebolt, 1985]{celeux1985sem}
Celeux, G. and Diebolt, J. (1985).
\newblock The {SEM} algorithm: a probabilistic teacher algorithm derived from
  the {EM} algorithm for the mixture problem.
\newblock {\em Computational Statistics Quarterly}, 2(1):73--82.

\bibitem[Dancik, 2013]{mlegp}
Dancik, G.~M. (2013).
\newblock {\em mlegp: Maximum Likelihood Estimates of Gaussian Processes}.
\newblock R package version 3.1.4.

\bibitem[Efron and Tibshirani, 1997]{efron1997improvements}
Efron, B. and Tibshirani, R. (1997).
\newblock Improvements on cross-validation: the 632+ bootstrap method.
\newblock {\em Journal of the American Statistical Association},
  92(438):548--560.

\bibitem[Fang et~al., 2005]{fang2005design}
Fang, K.-T., Li, R., and Sudjianto, A. (2005).
\newblock {\em Design and Modeling for Computer Experiments}.
\newblock CRC Press.

\bibitem[Fraley and Raftery, 2002]{fraley2002model}
Fraley, C. and Raftery, A.~E. (2002).
\newblock Model-based clustering, discriminant analysis, and density
  estimation.
\newblock {\em Journal of the American statistical Association},
  97(458):611--631.

\bibitem[Furrer et~al., 2006]{furrer2006covariance}
Furrer, R., Genton, M.~G., and Nychka, D. (2006).
\newblock Covariance tapering for interpolation of large spatial datasets.
\newblock {\em Journal of Computational and Graphical Statistics},
  15(3):502--523.

\bibitem[Gramacy, 2015]{gramacy2015lagp}
Gramacy, R.~B. (2015).
\newblock {laGP}: large-scale spatial modeling via local approximate {G}aussian
  processes in {R}.
\newblock {\em Journal of Statistical Software (available as a vignette in the
  laGP package)}.

\bibitem[Gramacy and Apley, 2015]{gramacy2015local}
Gramacy, R.~B. and Apley, D.~W. (2015).
\newblock Local {G}aussian process approximation for large computer
  experiments.
\newblock {\em Journal of Computational and Graphical Statistics},
  24(2):561--578.

\bibitem[Gramacy and Lee, 2008]{gramacy2008bayesian}
Gramacy, R.~B. and Lee, H. K.~H. (2008).
\newblock Bayesian treed {G}aussian process models with an application to
  computer modeling.
\newblock {\em Journal of the American Statistical Association},
  103(483):1119--1130.

\bibitem[Gramacy and Lee, 2009]{gramacy2009adaptive}
Gramacy, R.~B. and Lee, H. K.~H. (2009).
\newblock Adaptive design and analysis of supercomputer experiments.
\newblock {\em Technometrics}, 51(2):130--145.

\bibitem[Haaland and Qian, 2011]{haaland2011accurate}
Haaland, B. and Qian, P. Z.~G. (2011).
\newblock Accurate emulators for large-scale computer experiments.
\newblock {\em The Annals of Statistics}, 39(6):2974--3002.

\bibitem[Harville, 1998]{harville1998matrix}
Harville, D.~A. (1998).
\newblock {\em Matrix Algebra from a Statistician's Perspective}.
\newblock Springer, New York, NY.

\bibitem[Higdon et~al., 2002]{higdon2002space}
Higdon, D. et~al. (2002).
\newblock Space and space-time modeling using process convolutions.
\newblock {\em Quantitative Methods for Current Environmental Issues},
  3754:37--56.

\bibitem[Higdon et~al., 1999]{higdon1999non}
Higdon, D., Swall, J., and Kern, J. (1999).
\newblock Non-stationary spatial modeling.
\newblock {\em Bayesian Statistics}, 6(1):761--768.

\bibitem[Hwang et~al., 2018]{hwang2018}
Hwang, Y., Lu, S., and Kim, J.-K. (2018).
\newblock Bottom-up estimation and top-down prediction: Solar energy prediction
  combining information from multiple sources.
\newblock {\em Annals of Applied Statistics}, 12(4):2096--2120.

\bibitem[Kim et~al., 2005]{kim2005analyzing}
Kim, H.-M., Mallick, B.~K., and Holmes, C.~C. (2005).
\newblock Analyzing nonstationary spatial data using piecewise {G}aussian
  processes.
\newblock {\em Journal of the American Statistical Association},
  100(470):653--668.

\bibitem[Kohavi, 1995]{kohavi1995study}
Kohavi, R. (1995).
\newblock A study of cross-validation and bootstrap for accuracy estimation and
  model selection.
\newblock In {\em Proceedings of International Joint Conference on Artificial
  Intelligence}, pages 1137--1145.

\bibitem[Montagna and Tokdar, 2016]{montagna2016computer}
Montagna, S. and Tokdar, S.~T. (2016).
\newblock Computer emulation with nonstationary gaussian processes.
\newblock {\em SIAM/ASA Journal on Uncertainty Quantification}, 4(1):26--47.

\bibitem[Morris and Mitchell, 1995]{morris1995exploratory}
Morris, M.~D. and Mitchell, T.~J. (1995).
\newblock Exploratory designs for computational experiments.
\newblock {\em Journal of Statistical Planning and Inference}, 43(3):381--402.

\bibitem[Nguyen-Tuong and Peters, 2011]{nguyen2011model}
Nguyen-Tuong, D. and Peters, J. (2011).
\newblock Model learning for robot control: a survey.
\newblock {\em Cognitive processing}, 12(4):319--340.

\bibitem[Nguyen-Tuong et~al., 2009]{nguyen2009local}
Nguyen-Tuong, D., Peters, J., and Seeger, M. (2009).
\newblock Local gaussian process regression for real time online model
  learning.
\newblock In {\em Advances in Neural Information Processing Systems 21}, pages
  1193--1200.

\bibitem[Nielsen et~al., 2000]{nielsen2000stochastic}
Nielsen, S.~F. et~al. (2000).
\newblock The stochastic {EM algorithm: estimation and asymptotic results}.
\newblock {\em Bernoulli}, 6(3):457--489.

\bibitem[Nychka et~al., 2015]{nychka2014multi}
Nychka, D., Bandyopadhyay, S., Hammerling, D., Lindgren, F., and Sain, S.
  (2015).
\newblock A multi-resolution {G}aussian process model for the analysis of large
  spatial data sets.
\newblock {\em Journal of Computational and Graphical Statistics},
  24(2):579--599.

\bibitem[Paciorek and Schervish, 2006]{paciorek2006spatial}
Paciorek, C.~J. and Schervish, M.~J. (2006).
\newblock Spatial modelling using a new class of nonstationary covariance
  functions.
\newblock {\em Environmetrics}, 17(5):483--506.

\bibitem[Park and Apley, 2018]{park2018patchwork}
Park, C. and Apley, D. (2018).
\newblock Patchwork kriging for large-scale gaussian process regression.
\newblock {\em The Journal of Machine Learning Research}, 19(1):269--311.

\bibitem[Plumlee, 2014]{plumlee2014fast}
Plumlee, M. (2014).
\newblock Fast prediction of deterministic functions using sparse grid
  experimental designs.
\newblock {\em Journal of the American Statistical Association},
  109(508):1581--1591.

\bibitem[Plumlee and Apley, 2017]{plumlee2017lifted}
Plumlee, M. and Apley, D.~W. (2017).
\newblock Lifted brownian kriging models.
\newblock {\em Technometrics}, 59(2):165--177.

\bibitem[Qui{\~n}onero-Candela and Rasmussen, 2005]{quinonero2005unifying}
Qui{\~n}onero-Candela, J. and Rasmussen, C.~E. (2005).
\newblock A unifying view of sparse approximate {G}aussian process regression.
\newblock {\em Journal of Machine Learning Research}, 6(Dec):1939--1959.

\bibitem[{R Core Team}, 2015]{R2015}
{R Core Team} (2015).
\newblock {\em R: A Language and Environment for Statistical Computing}.
\newblock R Foundation for Statistical Computing, Vienna, Austria.

\bibitem[Rasmussen and Ghahramani, 2002]{rasmussen2002infinite}
Rasmussen, C.~E. and Ghahramani, Z. (2002).
\newblock Infinite mixtures of gaussian process experts.
\newblock In {\em Advances in neural information processing systems}, pages
  881--888.

\bibitem[Rasmussen and Williams, 2006]{rasmussen2006gaussian}
Rasmussen, C.~E. and Williams, C.~K. (2006).
\newblock {\em Gaussian processes for machine learning}, volume~1.
\newblock MIT press Cambridge.

\bibitem[{Revolution Analytics} and Weston, 2015]{pkgforeach}
{Revolution Analytics} and Weston, S. (2015).
\newblock {\em foreach: Provides Foreach Looping Construct for R}.
\newblock R package version 1.4.3.

\bibitem[Rogers et~al., 2009]{rogers2009ncep}
Rogers, E., DiMego, G., Black, T., Ek, M., Ferrier, B., Gayno, G., Janjic, Z.,
  Lin, Y., Pyle, M., Wong, V., et~al. (2009).
\newblock The ncep north american mesoscale modeling system: Recent changes and
  future plans.
\newblock In {\em 23rd Conference on Weather Analysis and Forecasting/19th
  Conference on Numerical Weather Prediction, Omaha, NE}.

\bibitem[Sang and Huang, 2012]{sang2012full}
Sang, H. and Huang, J.~Z. (2012).
\newblock A full scale approximation of covariance functions for large spatial
  data sets.
\newblock {\em Journal of the Royal Statistical Society: Series B},
  74(1):111--132.

\bibitem[Santner et~al., 2018]{santner2013design}
Santner, T.~J., Williams, B.~J., and Notz, W.~I. (2018).
\newblock {\em The Design and Analysis of Computer Experiments}.
\newblock Springer-Verlag New York, 2 edition.

\bibitem[Snelson and Ghahramani, 2006]{snelson2006sparse}
Snelson, E. and Ghahramani, Z. (2006).
\newblock Sparse {G}aussian processes using pseudo-inputs.
\newblock In {\em Advances in Neural Information Processing Systems}, pages
  1257--1264.

\bibitem[Stein, 2012]{stein2012interpolation}
Stein, M.~L. (2012).
\newblock {\em Interpolation of Spatial Data: Some Theory for Kriging}.
\newblock Springer Science \& Business Media.

\bibitem[Sun et~al., 2019a]{sun2017emulating}
Sun, F., Gramacy, R.~B., Haaland, B., Lawrence, E., and Walker, A. (2019a).
\newblock Emulating satellite drag from large simulation experiments.
\newblock {\em SIAM/ASA Journal on Uncertainty Quantification}.

\bibitem[Sun et~al., 2019b]{sun2018synthesizing}
Sun, F., Gramacy, R.~B., Haaland, B., Lu, S., and Hwang, Y. (2019b).
\newblock Synthesizing simulation and field data of solar irradiance.
\newblock {\em Statistical Analysis and Data Mining}, 12(4):311--324.

\bibitem[Sung, 2019]{sung2017mrfa}
Sung, C.-L. (2019).
\newblock {\em MRFA: Fitting and Predicting Large-Scale Nonlinear Regression
  Problems using Multi-Resolution Functional ANOVA (MRFA) Approach}.
\newblock R package version 0.4.

\bibitem[Sung et~al., 2020]{sung2017multi}
Sung, C.-L., Wang, W., Plumlee, M., and Haaland, B. (2020).
\newblock Multi-resolution functional {ANOVA} for large-scale, many-input
  computer experiments.
\newblock {\em Journal of the American Statistical Association},
  115(530):908--919.

\bibitem[Titsias, 2009]{titsias2009variational}
Titsias, M. (2009).
\newblock Variational learning of inducing variables in sparse {G}aussian
  processes.
\newblock In {\em Artificial Intelligence and Statistics}, pages 567--574.

\bibitem[Tresp, 2001]{tresp2001mixtures}
Tresp, V. (2001).
\newblock Mixtures of gaussian processes.
\newblock In {\em Advances in neural information processing systems}, pages
  654--660.

\bibitem[Xiong et~al., 2007]{xiong2007non}
Xiong, Y., Chen, W., Apley, D., and Ding, X. (2007).
\newblock A non-stationary covariance-based kriging method for metamodelling in
  engineering design.
\newblock {\em International Journal for Numerical Methods in Engineering},
  71(6):733--756.

\bibitem[Zachariassen et~al., 2003]{zachariassen2003review}
Zachariassen, J., Zeller, K.~F., Nikolov, N., and McClelland, T. (2003).
\newblock A review of the forest service remote automated weather station
  (raws) network.
\newblock {\em General Technical Report}.
\newblock No. RMRS-GTR-119.

\bibitem[Zhang et~al., 2019]{zhang2019learning}
Zhang, Y., Ghosh, S., Asher, I., Ling, Y., and Wang, L. (2019).
\newblock Learning uncertainty using clustering and local gaussian process
  regression.
\newblock In {\em AIAA Scitech 2019 Forum}, page 1730.

\end{thebibliography}


\begin{thebibliography}{}


\bibitem[Harville, 1998]{suppharville1998}
Harville, D. A. (1998).
\newblock {\em Matrix Algebra from a Statistician’s Perspective}.
\newblock Springer, New York.

\bibitem[Montagna and Tokdar, 2016]{suppmontagna2016}
Montagna, S. and Tokdar, S. T. (2016).  
\newblock Computer emulation with nonstationary Gaussian processes.
\newblock {\em SIAM/ASA Journal on Uncertainty Quantification}, 4(1):26--47.

\bibitem[Xiong et al., 2007]{suppxiong2007}
Xiong,  Y.,  Chen,  W.,  Apley,  D.,  and  Ding,  X.  (2007).  
\newblock A  non-stationarycovariance-based kriging method for metamodelling in engineering design.
\newblock {\em International Journal for Numerical Methods in Engineering}, 71(6):733--756.


\end{thebibliography}
}

\def\spacingset#1{\renewcommand{\baselinestretch}%
{#1}\small\normalsize} \spacingset{1.3}

\newpage
\setcounter{page}{1}
\bigskip
\bigskip
\bigskip
\begin{center}
{\Large\bf Supplementary Materials for ``A Clustered Gaussian Process Model for Computer Experiments''}
\end{center}
\medskip

\setcounter{section}{0}
\setcounter{equation}{0}
\setcounter{figure}{0}
\def\theequation{S\arabic{section}.\arabic{equation}}
\def\thesection{S\arabic{section}}
\def\thefigure{S\arabic{figure}}
\def\thetable{S\arabic{table}}
\section{Proof of Proposition \ref{prop1}}\label{append:proof}
For notational convention, denote $\Sigma_j=\Phi_{\gamma_j}(X_{\mathcal{P}_j\setminus\{i\}},X_{\mathcal{P}_j\setminus\{i\}})$ and $W_j=Y_{\mathcal{P}_j\setminus\{i\}}-\mu_j(X_{\mathcal{P}_j\setminus\{i\}})$ for $j=1,\ldots,K$. Then, for any $j\neq k$,
\begin{align}\label{eq:appendA1}
f_j(Y_{\mathcal{P}_j\setminus\{i\}}|X_{\mathcal{P}_j\setminus\{i\}};\theta_j)=\frac{1}{\sqrt{2\pi\det(\Sigma_j)}}\exp\left\{-\frac{1}{2} W_j^T\Sigma^{-1}_jW_j\right\},
\end{align}
by the fact that $f_j$ is the probability density function of a multivariate normal distribution with parameters $\theta_j=(\mu_j(\cdot),\sigma_j^2,\gamma_j)$. For $j=k$, by partitioned matrix inverse and determinant formulas,
\begin{align}\label{eq:appendA2}
&f_k(Y_{\mathcal{P}_k\cup\{i\}}|X_{\mathcal{P}_k\cup\{i\}})\nonumber\\
=&\frac{1}{\sqrt{2\pi\det\left(\left[\begin{array}{cc}\Sigma_k& r^T_{i,-i}\\r_{i,-i} & \sigma^2_k\end{array}\right]\right)}}\exp\left\{-\frac{1}{2} \left[\begin{array}{c}W_k\\y_i-\mu_k(x_i)\end{array}\right]^T\left[\begin{array}{cc}\Sigma_k& r^T_{i,-i}\\r_{i,-i} & \sigma^2_k\end{array}\right]^{-1}\left[\begin{array}{c}W_k\\ y_i-\mu_k(x_i)\end{array}\right]\right\}\nonumber\\
=&f_k(Y_{\mathcal{P}_k\setminus\{i\}}|X_{\mathcal{P}_k\setminus\{i\}})\times \frac{1}{\sqrt{(\sigma^*_k)^2}}\exp\left\{-\frac{1}{2}(y_i-\mu^*_k)^2/(\sigma^*_k)^2\right\},
\end{align}
where $r_{i,-i}=\Phi_{\gamma_k}(x_i,X_{P_k\setminus\{i\}})$, $\mu^*_k=\mu_k(x_i)+r_{i,-i}\Sigma_k^{-1}W_k$ and
$(\sigma_k^*)^2=\sigma_k^2(1-r_{i,-i}\Sigma_k^{-1}r_{i,-i}^T).$

Therefore, combining \eqref{GibbsAssignment}, \eqref{eq:appendA1} and \eqref{eq:appendA2}, 
\begin{align*}
f(z_i=k|X,Y,Z_{-i})\propto &f_k(Y_{\mathcal{P}_k\cup\{i\}}|X_{\mathcal{P}_k\cup\{i\}};\theta_k)\prod_{j\ne k}f_{j}(Y_{\mathcal{P}_{j}\setminus\{i\}}|X_{\mathcal{P}_{j}\setminus\{i\}};\theta_j)g_k(x_i;\varphi_k)\\
=&\prod^K_{k=1}f_k(Y_{\mathcal{P}_k\setminus\{i\}}|X_{\mathcal{P}_k\setminus\{i\}};\theta_k)\exp\left\{-\frac{1}{2}(y_i-\mu^*_k)^2/(\sigma^*_k)^2\right\}g_k(x_i;\varphi_k)\\
\propto&\phi((y_i-\mu^*_k)/\sigma^*_k)g_k(x_i;\varphi_k).
\end{align*}

\section{Efficient Update for the Stochastic E-step}\label{app:updatematrixinverse}
In this section, partitioned matrix inverse formula is introduced to efficiently update the mean and variance of \eqref{eq:GibbsAssignment2} when looping through observation $i$ in the stochastic E-step. Suppose that the current assignment of observation $i$ is $z(x_i)=k$ but the new assignment of it is $z(x_i)=s$ where $s\neq k$, then the sets $\mathcal{P}_k$ and $\mathcal{P}_s$ will be updated, that is, $\mathcal{P}'_k\leftarrow\mathcal{P}_k\setminus\{i\}$ and $\mathcal{P}'_s\leftarrow\mathcal{P}_s\cup\{i\}$. The matrix inverses of $\Phi_{\gamma_k}(X_{\mathcal{P}'_k},X_{\mathcal{P}'_k})$ and $\Phi_{\gamma_s}(X_{\mathcal{P}'_s},X_{\mathcal{P}'_s})$ can be updated accordingly via partitioned matrix inverse formula as follows. Let $U\in\mathbb{R}^{n_k\times 2}$, where $n_k$ is the number of observations in the set $\mathcal{P}_k$, and $U_{i,1}=1$ and $U_{-i,2}=\Phi_{\gamma_k}(X_{P_k},x_i)$, otherwise $U_{i,j}=0$. For notational simplicity, denote $A=\Phi_{\gamma_k}(X_{\mathcal{P}_k},X_{\mathcal{P}_k})^{-1}$. Then, by the Woodbury formula \citep{suppharville1998}, the matrix inverses of $\Phi_{\gamma_k}(X_{\mathcal{P}'_k},X_{\mathcal{P}'_k})$ can be updated by
$$\Phi_{\gamma_k}(X_{\mathcal{P}'_k},X_{\mathcal{P}'_k})^{-1}=\left(A+AU(I_{2}-U^TAU)^{-1}U^TA\right)_{-i,-i},$$
where $I_{2}\in\mathbb{R}^2$ is a diagonal matrix.

Let $V=\Phi_{\gamma_s}(X_{P_s},x_i)\in\mathbb{R}^{n_s\times 1}$ and denote $B=\Phi_{\gamma_s}(X_{\mathcal{P}_s},X_{\mathcal{P}_s})^{-1}$. Then, by the partitioned matrix inverse formula \citep{suppharville1998}, the matrix inverses of $\Phi_{\gamma_s}(X_{\mathcal{P}'_s},X_{\mathcal{P}'_s})$ can be updated by
\begin{align*}
    \left(\Phi_{\gamma_k}(X_{\mathcal{P}'_s},X_{\mathcal{P}'_s})^{-1}\right)_{i,i}&=1/(1-V^TBV),\\
    \left(\Phi_{\gamma_k}(X_{\mathcal{P}'_s},X_{\mathcal{P}'_s})^{-1}\right)_{i,-i}&=-V^TB/(1-V^TBV)=\left(\Phi_{\gamma_k}(X_{\mathcal{P}'_s},X_{\mathcal{P}'_s})^{-1}\right)^T_{-i,i},\\
     \left(\Phi_{\gamma_k}(X_{\mathcal{P}'_s},X_{\mathcal{P}'_s})^{-1}\right)_{-i,-i}&=B+BVV^TB/(1-V^TBV).
\end{align*}

\section{Stochastic EM algorithm for clustered Gaussian process}\label{alg:clusteringalgorithm}
\begin{tabbing}
\enspace \textbf{Initialization}: \\
   \qquad Set $K$ clusters with random memberships $\{z(x_i)\}^n_{i=1}$\\
   \qquad Set $\mathcal{P}_k\leftarrow\{i:z(x_i)=k\}$ for each $k$\\
   \qquad Set initial parameters $\theta_k=\{\mu_k(\cdot),\sigma^2_k,\gamma_k\}$ and $\varphi_k$ for $k=1,\ldots,K$\\
   \enspace \textbf{Stochastic E-Step}: \\
   \qquad For $i=1$ to $i=n$,\\
   \qquad\qquad For $k=1$ to $K$ do parallel,\\
   \qquad\qquad\qquad $\mu^*_k\leftarrow \mu_k(x_i)+\Phi_{\gamma_k}(x_i,X_{\mathcal{P}_k\setminus\{i\}})\Phi_{\gamma_k}(X_{\mathcal{P}_k\setminus\{i\}},X_{\mathcal{P}_k\setminus\{i\}})^{-1}\left(Y_{\mathcal{P}_k\setminus\{i\}}-\mu_k(X_{\mathcal{P}_k\setminus\{i\}})\right)$\\
   \qquad\qquad\qquad $(\sigma_k^*)^2\leftarrow\sigma_k^2\left(1-\Phi_{\gamma_k}(x_i,X_{\mathcal{P}_k\setminus\{i\}})\Phi_{\gamma_k}(X_{\mathcal{P}_k\setminus\{i\}},X_{\mathcal{P}_k\setminus\{i\}})^{-1}\Phi_{\gamma_k}(X_{\mathcal{P}_k\setminus\{i\}},x_i)\right)$\\
   \qquad\qquad\qquad  $p_{ik}\leftarrow \frac{\phi((y_i-\mu^*_k)/\sigma_k^*)g_k(x_i;\varphi_k)}{\sum^K_{k=1}\phi((y_i-\mu^*_k)/\sigma_k^*)g_k(x_i;\varphi_k)}$\\
\qquad \qquad Draw $z$ from a random multinomial cluster assignment with probabilities $(p_{i1},\ldots,p_{iK})$\\
\qquad\qquad Update $z(x_i)\leftarrow z$\\
\qquad\qquad Update $\mathcal{P}_k\leftarrow\{i:z(x_i)=k\}$ for each $k$\\
\enspace \textbf{M-Step}: \\
\qquad For $k=1$ to $K$ do parallel,\\
\qquad\qquad Update $\theta_k\leftarrow\arg\max_{\theta_k}\log f_k(Y_{\mathcal{P}_k}|X_{\mathcal{P}_k};\theta_k)\pi(\theta_k)$\\
\qquad Update $\{\varphi_k\}^K_{k=1}\leftarrow\arg\max_{\boldsymbol{\varphi}}\sum_{k=1}^K\left(\sum_{i\in\mathcal{P}_k}\log g_k(x_i;\varphi_k)+\log\pi(\varphi_k)\right)$ \\
\enspace \textbf{Iteration}: Iterate stochastic E-step and M-step until some stopping rule is met.\\ 
\enspace \textbf{Output} $\{z(x_i)\}^n_{i=1}$, $\{\theta_k,\varphi_k\}^K_{k=1}$

\end{tabbing}

\section{One-dimensional examples}\label{supp:1d_example}
\begin{figure}[ht]
\centering
\includegraphics[width=0.8\linewidth]{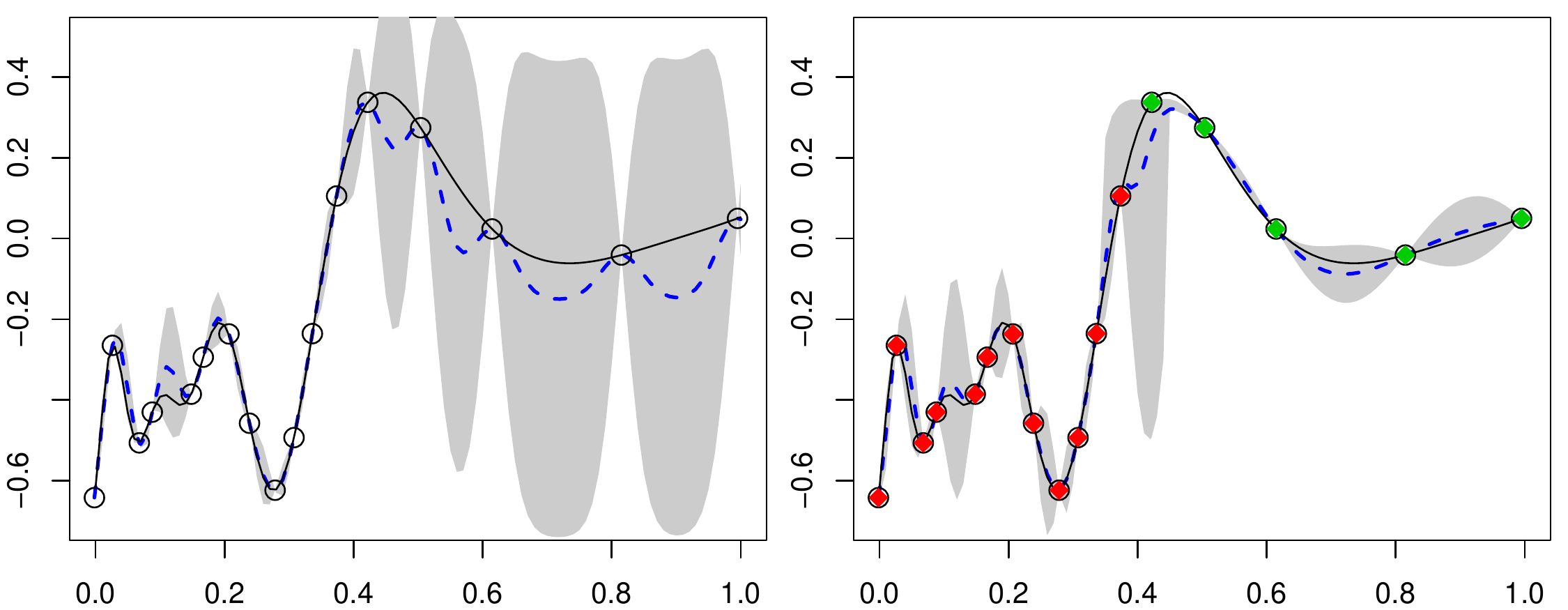}
\includegraphics[width=0.8\linewidth]{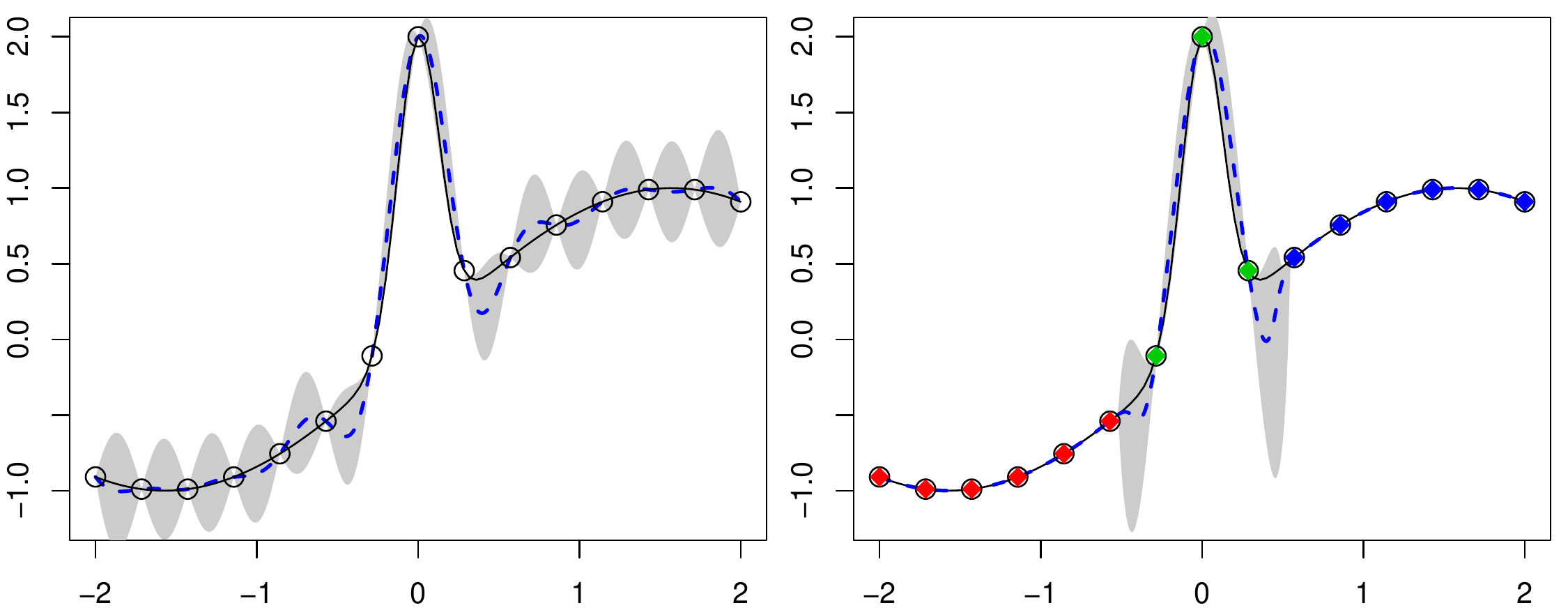}
\caption{One-dimensional synthetic data from (top)\cite{suppxiong2007} and (bottom) \cite{suppmontagna2016}. The left and right panels illustrate predictors by a stationary GP and a clustered GP, respectively. Black line is the true function, black circles are input locations, and blue dotted lines are the predictors, with the gray shaded region providing a pointwise 95\% confidence band. Red, green, and blue dots in the right panels represent different clusters.}
\label{fig:example_synthetic2insupp}
\end{figure}

Two more one-dimensional examples of Section \ref{sec:onedimexample} are presented here. Consider another example from \cite{suppxiong2007}, where the true function is 
\[
f(x)=\sin(30(x-0.9)^4)\cos(2(x-0.9))+(x-0.9)/2
\]
and 17 unequally spaced points from $[0,1]$ are chosen to evaluate. Similarly, the top panels of Figure \ref{fig:example_synthetic2insupp} show that the clustered GP (right) outperforms the stationary GP (left) in terms of prediction accuracy and uncertain quantification. The two clusters are separated at location around $x=0.40$. In particular, the predictor in the region $[0.42,1.00]$ has better prediction accuracy with much smaller prediction uncertainty. The same argument applies to this example: the constant mean and variance assumptions are violated in this function so the stationary GP results in the erratic prediction in the region $[0.42,1.00]$.

Lastly, consider the inhomogeneous smooth function in \cite{suppmontagna2016},
\[
f(x)=\sin(x)+2\exp(-30x^2),
\]
and 15 unequally spaced points from $[-2,2]$ are chosen to evaluate. The bottom panels of Figure \ref{fig:example_synthetic2insupp} demonstrates a stationary GP (left), where the prediction mean curve has large oscillations with confidence intervals except the tall peak in the middle. The is due to the rippling effect of the discovery of a tall peak, and \cite{suppmontagna2016} called the phenomenon a \textit{spline tension} effect in the predictor form. The clustered GP (right) overcomes the issue by separating the input locations into three clusters and fits a stationary GP in each cluster. The result shows that the prediction mean curve quite matches the true curve with a narrower confidence band. 

\section{Supporting Tables and Figures in Sections \ref{sec:numericalstudy} and \ref{sec:realdata}}\label{append:casestudyfigure}
The figures and tables that present the results in Sections \ref{sec:numericalstudy} and \ref{sec:realdata} are provided in this section.

\begin{table}[h!]
\centering
\caption{Borehole function example with $n$ training samples $n_{\rm test}=10,000$ testing locations.}
\begin{tabular}{
c|c|cccc}
\toprule 
Method & $n$ & Fitting  & Prediction & RMSE \\
&& Time (sec.) & Time (sec.) & \\
\midrule
\texttt{mlegp} & 1,000 & 5204 & 24  & 1.0902\\
\midrule
\multirow{3}{*}{\texttt{laGP}} & 1,000 & - & 153 & 1.1806 \\
 & 10,000& - & 137 & 0.4149 \\
 & 100,000 & - & 144 &  0.1617\\
 \midrule
\multirow{3}{*}{\texttt{MRFA}} & 1,000 & 116 &  17 & 0.4668 \\
 & 10,000& 723 & 16 & 0.0844 \\
 & 100,000 & 6789 & 18 & 0.0827\\
\midrule
\multirow{3}{*}{\texttt{clusterGP}} & 1,000 & 255 & 9 & 0.1124 \\
 & 10,000& 2950 & 55 & 0.0689 \\
 & 100,000 & 28434 & 535 & 0.0523 \\
\bottomrule
\end{tabular}
\label{tab:borehole}
\end{table}

\begin{figure}[ht]
\centering
\includegraphics[width=0.9\linewidth]{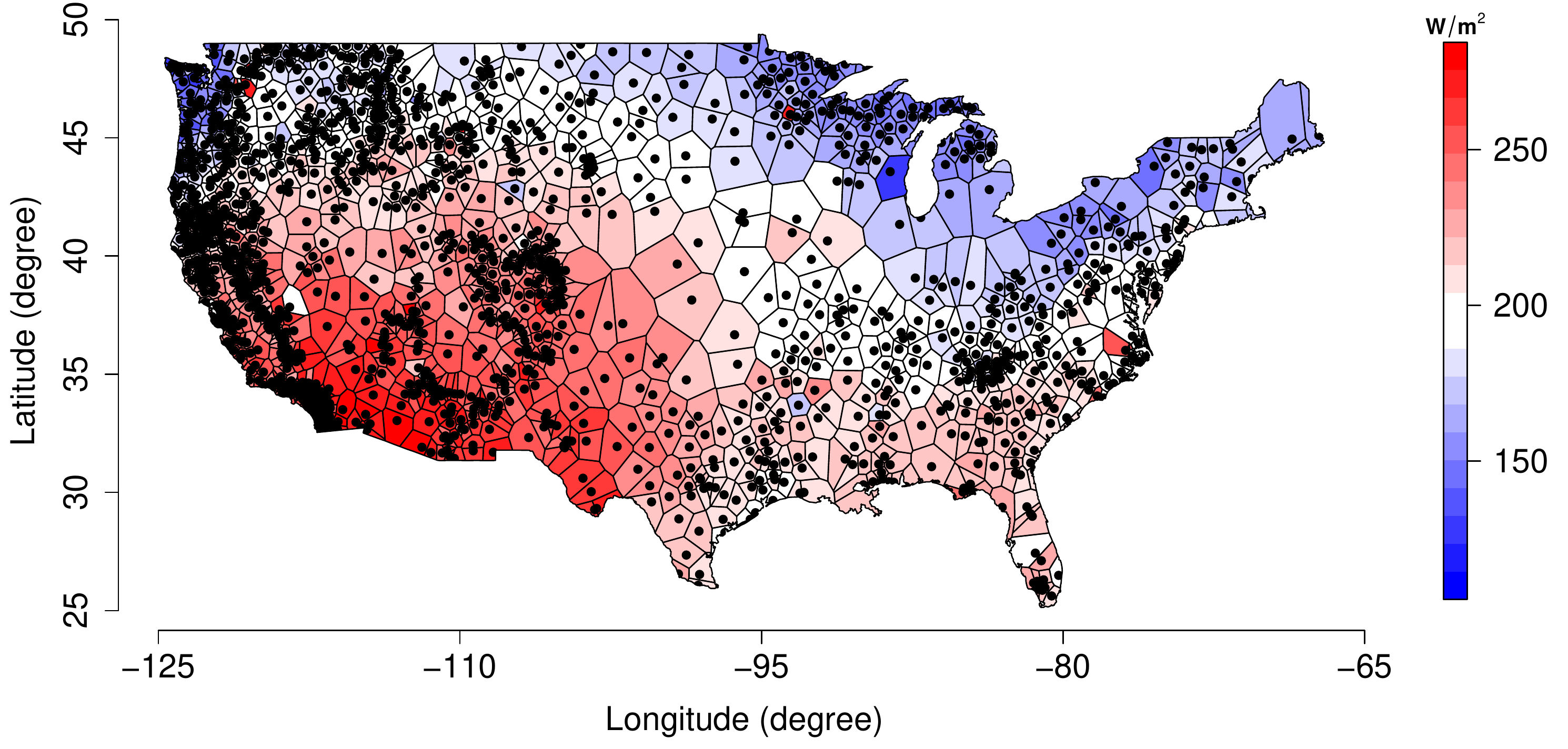}
\caption{Solar irradiance simulation from the North American Mesoscale Forecast System (NAM). The black dots are the Remote Automatic Weather Station (RAWS) measurement sites in the contiguous United States from which the NAM simulations are extracted. The regional colors represent the solar irradiance in the subfield of a particular measurement site.}
\label{fig:solarstation}
\end{figure}

\begin{figure}[h!]
\centering
\includegraphics[width=0.85\linewidth]{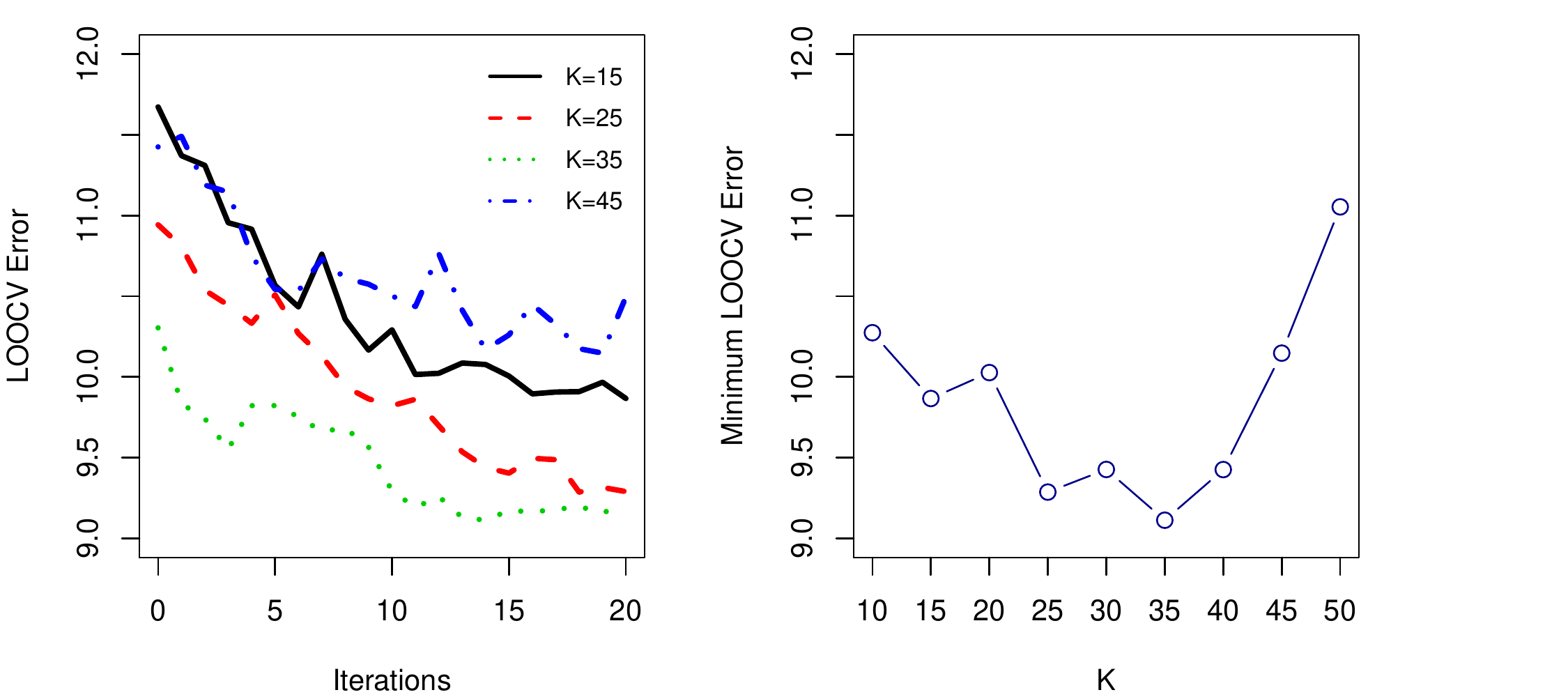}
\caption{The LOOCV RMSEs with $K=15,25,35$ and $45$ during the 20 iteration of the stochastic EM algorithm (left), and the minimum LOOCV RMSEs of $K=15, 20, 25, 30, 35, 40, 45, 50$ (right).}
\label{fig:determineK}
\end{figure}

\begin{figure}[h!]
\centering
\includegraphics[width=0.9\linewidth]{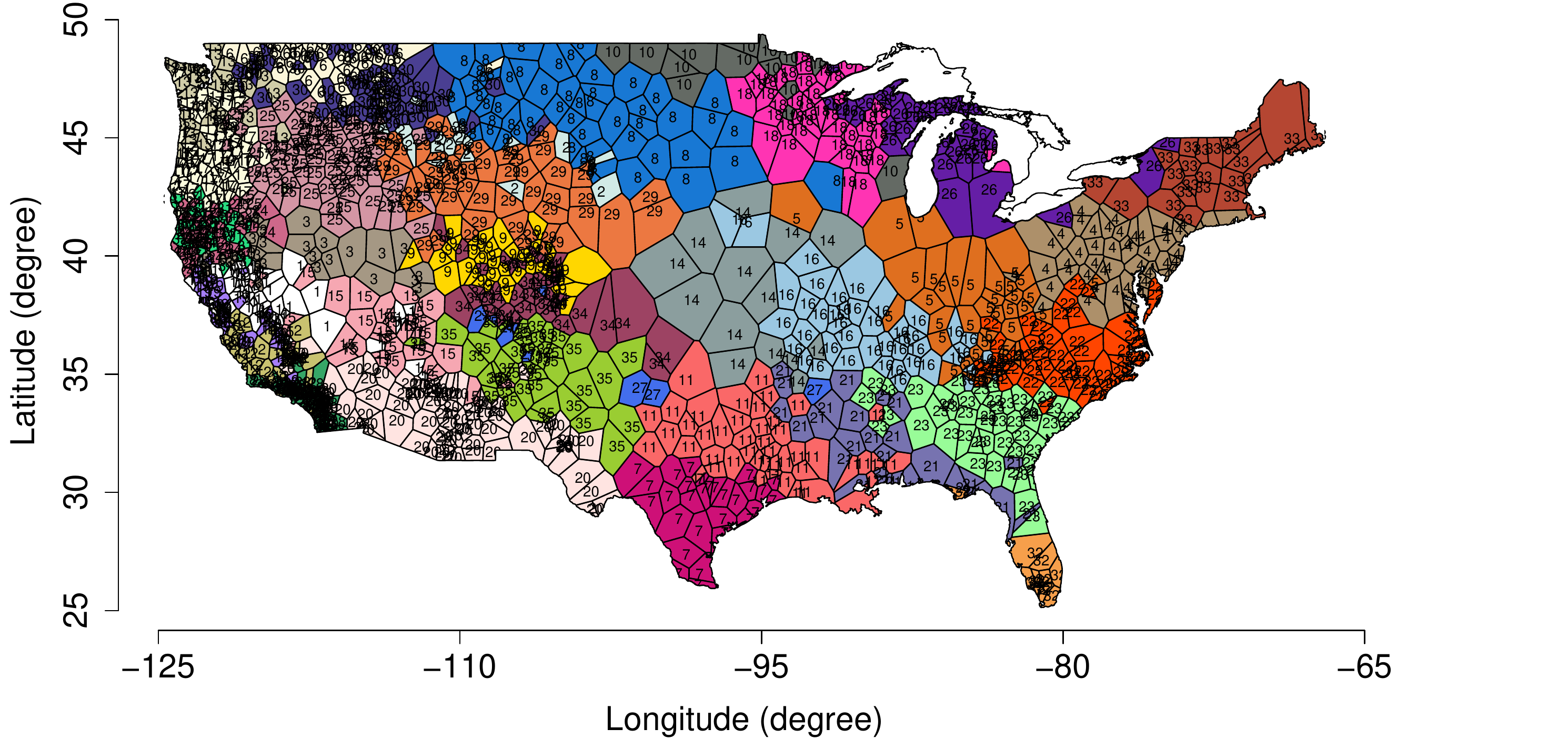}
\caption{Visualization of the cluster assignments with $K=35$.}
\label{fig:clustering}
\end{figure}

\begin{figure}[ht]
\centering
\includegraphics[width=0.9\linewidth]{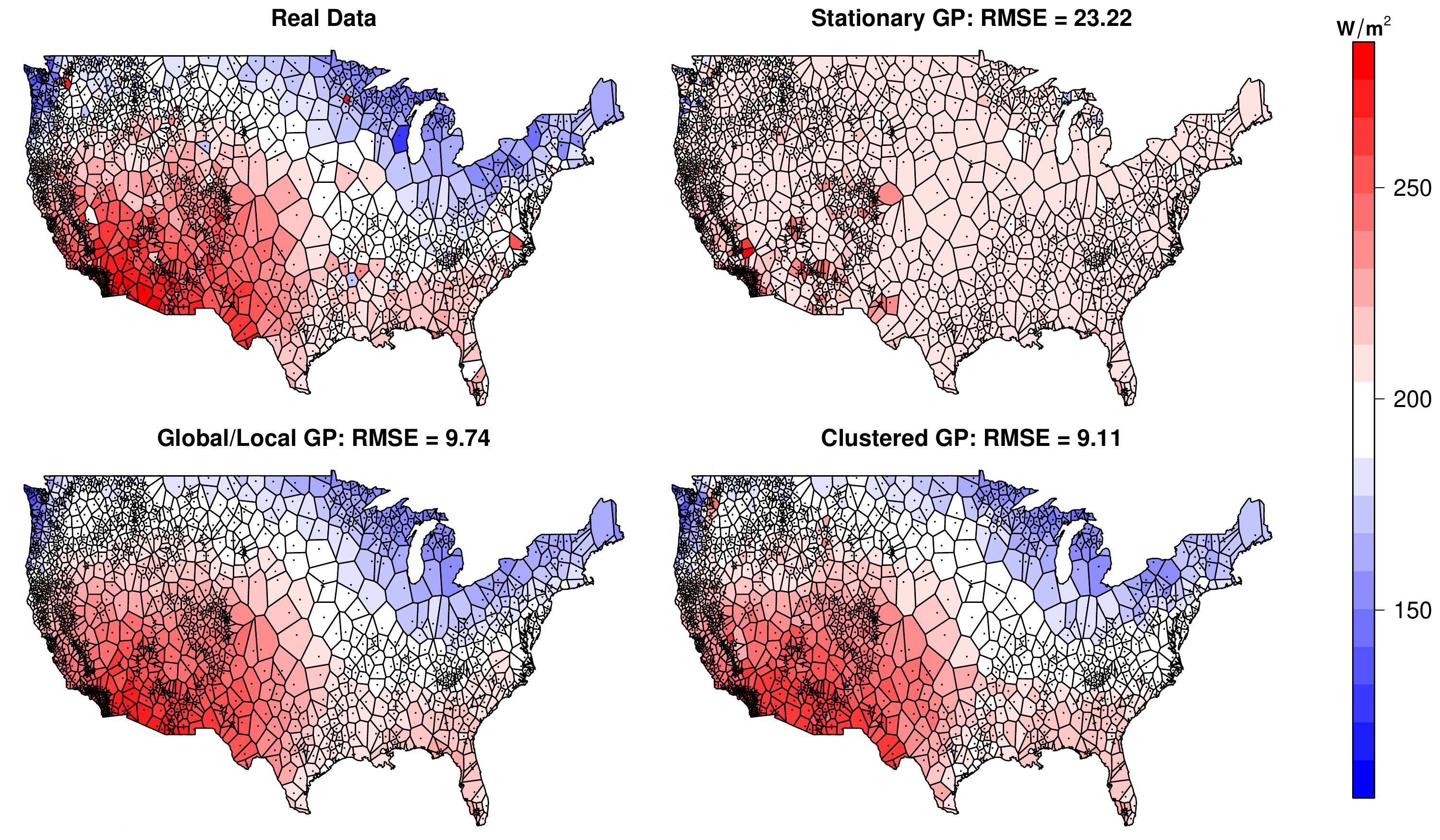}
\caption{Comparison of solar irradiance predictions. The true solar irradiance (top left), and the LOOCV predictions of a stationary GP (top right), a multi-resolution global/local GP (bottom left), and a clustered GP with $K=35$ (bottom right) are presented, along with their corresponding LOOCV RMSEs in the figure titles.}
\label{fig:predictioncomparison}
\end{figure}

\end{document}